\documentclass[article]{aa}  

\usepackage{graphicx}
\usepackage{txfonts}
\usepackage{lscape}
\usepackage{booktabs}
\usepackage{threeparttablex}
\usepackage{wasysym}
\usepackage{arydshln}
\usepackage[bookmarks=false,colorlinks=true,linkcolor=black,citecolor=cyan,filecolor=black,urlcolor=cyan]{hyperref}
\usepackage{array}

\newcolumntype{C}[1]{>{\centering\arraybackslash}p{#1}}

\begin{document} 
	
	
	\title{O Corona, where art thou? eROSITA's view of UV-optical-IR variability-selected massive black holes in low-mass galaxies}
	  \subtitle{}
	
	\author{R. Arcodia\thanks{NASA Einstein fellow}
		\inst{1,2},
        A. Merloni
		\inst{1},
        J. Comparat
		\inst{1},
        T. Dwelly
		\inst{1},
        R. Seppi
		\inst{1},
        Y. Zhang
		\inst{1},
        J. Buchner
		\inst{1},
        A. Georgakakis
		\inst{3},
        F. Haberl
		\inst{1},
        Z. Igo
		\inst{1},
        E. Kyritsis
        \inst{4,5},
        T. Liu
		\inst{1},
        K. Nandra
		\inst{1},
        Q. Ni
		\inst{1},
        G. Ponti
		\inst{6,1},
        M. Salvato
		\inst{1},
        C. Ward
        \inst{7},
        J. Wolf
		\inst{1,8,9},
        A. Zezas
        \inst{4,5}
	}
	
	\institute{Max-Planck-Institut f\"ur extraterrestrische Physik (MPE), Gie{\ss}enbachstra{\ss}e 1, 85748 Garching bei M\"unchen, Germany\\
		\email{rarcodia@mit.edu}
		\and
		MIT Kavli Institute for Astrophysics and Space Research, 70 Vassar Street, Cambridge, MA 02139, USA
		\and
		Institute for Astronomy \& Astrophysics, National Observatory of Athens, V. Paulou \& I. Metaxa 11532, Greece
        \and
		Department of Physics, University of Crete, Voutes University campus, 70013 Heraklion, Greece
        \and
        Institute of Astrophysics, Foundation for Research and Technology-Hellas, N. Plastira 100, Vassilika Vouton, 71110 Heraklion, Greece
        \and
        INAF-Osservatorio Astronomico di Brera, Via Bianchi 46, 23807 Merate, LC, Italy
        \and
        Department of Astrophysical Sciences, Princeton University, Princeton, NJ 08544, USA
        \and
        Exzellenzcluster ORIGINS, Boltzmannstr. 2, 85748 Garching bei M\"unchen, Germany
        \and
        Max-Planck-Institut f\"ur Astronomie, Königstuhl 17, 69117 Heidelberg, Germany
		}
	
	\date{Received ; accepted }
	
	\abstract{Finding massive black holes (MBHs, $M_{BH}\approx10^4-10^7 M_{\astrosun}$) in the nuclei of low-mass galaxies ($M_{*}\lessapprox10^{10} M_{\astrosun}$) is crucial to constrain seeding and growth of black holes over cosmic time, but it is particularly challenging due to their low accretion luminosities. Variability selection via long-term photometric ultraviolet, optical, or infrared (UVOIR) light curves has proved effective and identifies lower-Eddington ratios compared to broad and narrow optical spectral lines searches. In the inefficient accretion regime, X-ray and radio searches are effective, but they have been limited to small samples. Therefore, differences between selection techniques have remained uncertain. Here, we present the first large systematic investigation of the X-ray properties of a sample of known MBH candidates in dwarf galaxies. We extracted X-ray photometry and spectra of a sample of $\sim200$ UVOIR variability-selected MBHs and significantly detected 17 of them in the deepest available \emph{SRG}/eROSITA image, of which four are newly discovered X-ray sources and two are new secure MBHs. This implies that tens to hundreds of LSST MBHs will have \emph{SRG}/eROSITA counterparts, depending on the seeding model adopted. Surprisingly, the stacked X-ray images of the many non-detected MBHs are incompatible with standard disk-corona relations, typical of active galactic nuclei, inferred from both the optical and radio fluxes. They are instead compatible with the X-ray emission predicted for normal galaxies. After careful consideration of potential biases, we identified that this X-ray weakness needs a physical origin. A possibility is that a canonical X-ray corona might be lacking in the majority of this population of UVOIR-variability selected low-mass galaxies or that unusual accretion modes and spectral energy distributions are in place for MBHs in dwarf galaxies. This result reveals the potential for severe biases in occupation fractions derived from data from only one waveband combined with SEDs and scaling relations of more massive black holes and galaxies.
    }
	
	
	\titlerunning{O Corona, where art thou?}
	\authorrunning{R. Arcodia et al.} 
	\maketitle

	%
		
	\section{Introduction}  
	
	It is hotly debated to what extent the nuclei of low-mass galaxies (i.e., stellar masses $M_{*}\lessapprox10^{10} M_{\astrosun}$) are populated by massive black holes (MBHs), a fairly loose term naming masses intermediate in between stellar and super-massive (used here for the range $M_{BH}\approx10^4-10^7 M_{\astrosun}$; e.g., see \citealp{Greene+2020:imbhs} and references therein). An in-depth understanding of this population of nearby low-mass nuclei is fundamental in relation to the first early Universe galaxies which they closely resemble. However, predictions on this local population from theoretical grounds require assumptions on seeding origin and growth \citep[e.g., see][]{Bellovary+2019:MBHs,Pacucci+2021:MBHs,Haidar+2022:Occ,Beckmann+2023:newhorizonsimul}. Instead, from observational grounds we are fundamentally limited by the fraction of massive black holes which, even if they exist, are effectively active and luminous enough to be discernible from the host galaxy's emission at any wavelength \citep[e.g.,][]{Greene+2020:imbhs,Reines2022:MBHs}.  
	
	The main channel used so far to systematically select MBHs is optical spectroscopy. The brightest end (in terms of the Eddington-normalized luminosity, $L/L_{edd}$) can be unveiled through virial mass estimates inferred from broad lines \citep[e.g.,][]{Greene+2004:sdssI,Greene+2007:sdssIV,Chilingarian+2018:imbhs, Salehirad+2022:GAMAbl}, yielding $\sim500$ MBHs to date \citep{Greene+2020:imbhs}. Understandably, this selection merely scratches the surface of the population of nuclear MBHs in low-mass galaxies, as only a very small fraction of galactic nuclei \citep[$\lessapprox1\%$; e.g.,][]{Bongiorno+2012:prob,Georgakakis+2017:prob} are expected to be in the range of the required $L/L_{edd}$ to show strong broad lines, even more so for low-mass galaxies \citep{Aird+2012:prob}. Narrow-line-based classifications \citep{Baldwin+1981:bpt} may find low-mass galaxies with evidence of hard ionization from a nuclear source \citep[e.g.,][]{Barth+2008:SDSSbpt,Reines+2013:dwBPT,Moran+2014:BPT,Sartori+2015:bpt} at lower $L/L_{edd}$. Of course, the fainter these active MBHs are, the more they get inevitably hidden by the host galaxy's stellar emission and their signatures become hardly distinguishable from those of star-forming galaxies \citep[e.g.,][]{Cann+2019:limit}. Spatially resolving emission from the nucleus helps \citep{Mezcua+2020:MANGA_lowM}, although this approach is limited by angular resolution and therefore distance. Furthermore, a small fraction of nuclear MBHs can be unveiled through bright transient accretion events, for instance tidal disruptions of stars \citep[e.g.,][]{Donato+2014:imbhtde,He+2021:imbhtde,Angus+2022:fastrise} and, lately, the puzzling quasi-periodic eruptions \citep{Miniutti+2019:qpe1,Giustini+2020:qpe2,Arcodia+2021:eroqpes,Chakraborty+2021:qpe5cand}, although this channel is limited by the low volumetric rates of these events (\citealp{Vanvelzen+2020:rates,Angus+2022:fastrise}; Arcodia et al., in prep.).
	
    An alternative and promising way forward is given by the growing number of high-cadence photometric surveys, which allow for the selection of MBHs through optical, ultraviolet (UV), and infrared (IR) variability \citep[UVOIR variability hereafter;][]{Shaya+2015:kepler,Baldassare+2018:SDSS,Baldassare+2020:PTF,Martinez-Palomera2020:sibling,Kimura+2020:HSC,Elmer+2020:nir,Secrest+2020:Wise,Ward+2022:ZTF_Wise,Burke+2022:DES,Shin+2022:intranight,Wasleske+2022:Galex,Ward+2022:ZTF_Wise}. The goal of this method is to find evidence of low-level photometric variability through difference imaging analysis, indicative of nuclear point-like sources embedded in their extended host galaxies. Most of these studies compare light curves to a damped random walk model, which is usually an empirical indicator of accretion variability in active galactic nuclei \citep[AGN; e.g.,][]{Kelly+2009:DRW,Butler+2011:sigma}. This method was shown to yield a larger detection rate of MBH candidates below $M_{*}\sim10^{10} M_{\astrosun}$, compared to broad and narrow line selection techniques \citep[e.g.,][]{Baldassare+2018:SDSS,Baldassare+2020:PTF}. 
	The radio and X-ray band are more suitable to find nuclear sources in low-mass galaxies, as they have a higher nuclear-to-host contrast \citep{Merloni2016:contrast}. Therefore, a dedicated follow-up with deep X-ray and radio observations can serve to strengthen these candidates further \citep[e.g,][]{Reines+2011:Hen,Latimer+2019:chandraVLA_BCG,Latimer2021:ChHST,Graham+2021:Xvirgo,Davis+2022:radioHSC}, as well as performing matches with current X-ray archives \citep{Schramm+2013:chandra,Lemons+2015:chandra,Pardo+2016:chandra,Mezcua+2018:cosmos,Birchall+2020:xmmsdss,Latimer+2021:efeds,Bykov+2023:dwarves}. However, the former method is not a viable option for all of the known low-mass galaxies in the sky and the latter has been naturally limited in sky area so far. This is where the extended ROentgen Survey with an Imaging Telescope Array \citep[eROSITA;][]{Predehl+2021:eROSITA} aboard the Spectrum-Roentgen-Gamma observatory \citep[\emph{SRG};][]{Sunyaev+2021:SRG} comes into play with its all-sky survey capabilities, complementing existing deep-exposure and narrow-field datasets \citep[e.g.,][for a recent showcase]{Bykov+2023:dwarves}.
	
	Here, we focus on MBHs selected through UVOIR variability (Sect.~\ref{sec:sample}), which has the advantage of providing a sample with occupation and an active fraction of one. Therefore, for this work we used MBHs and accreting central black holes in low-mass galaxies interchangeably. We systematically extracted X-ray properties from the eROSITA all-sky survey data (Sect.~\ref{sec:xray}). The primary goal was to obtain their X-ray detection fraction (Sect.~\ref{sec:results}), providing a top tier of UVOIR-variable X-ray-detected MBHs in low-mass galaxies for future deeper multiwavelength studies (Sect.~\ref{sec:discussion_det}), and to calibrate how single-band searches for MBHs compare (Sect.~\ref{sec:discussion_xweak}). This work will also serve as a pilot study to understand the connection between variability selection methods and eROSITA X-ray data to exploit future synergies with the Vera C. Rubin Observatory Legacy Survey of Space and Time \citep[LSST;][]{Ivezic+2019:LSST}. 

    \section{Sample selection}
    \label{sec:sample}

    \begin{figure}[t]
    \centering
    \includegraphics[width=0.83\columnwidth]{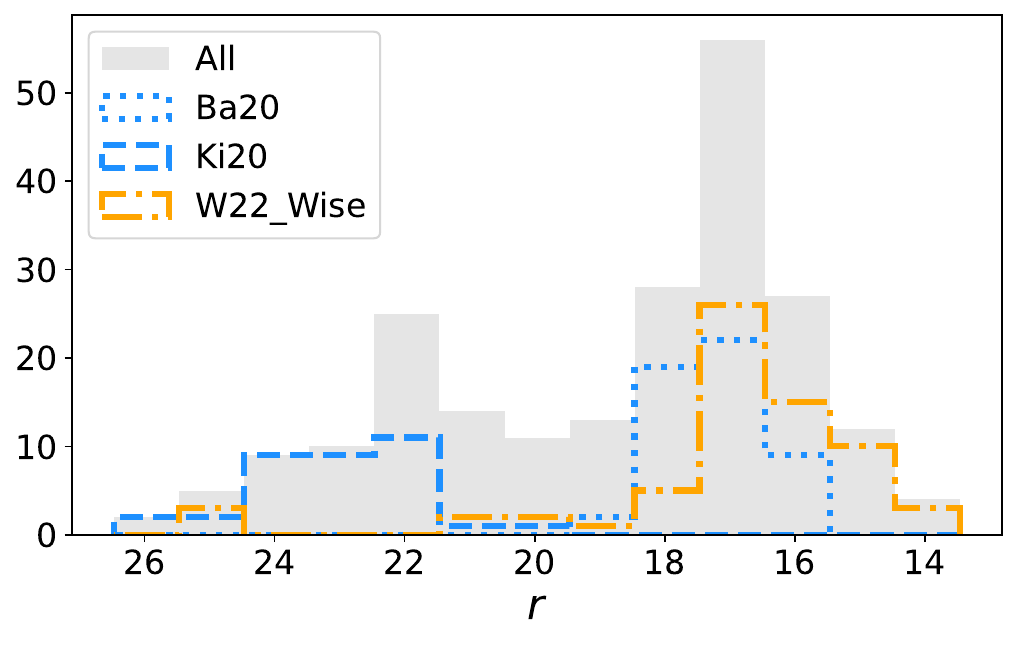}
    \includegraphics[width=0.83\columnwidth]{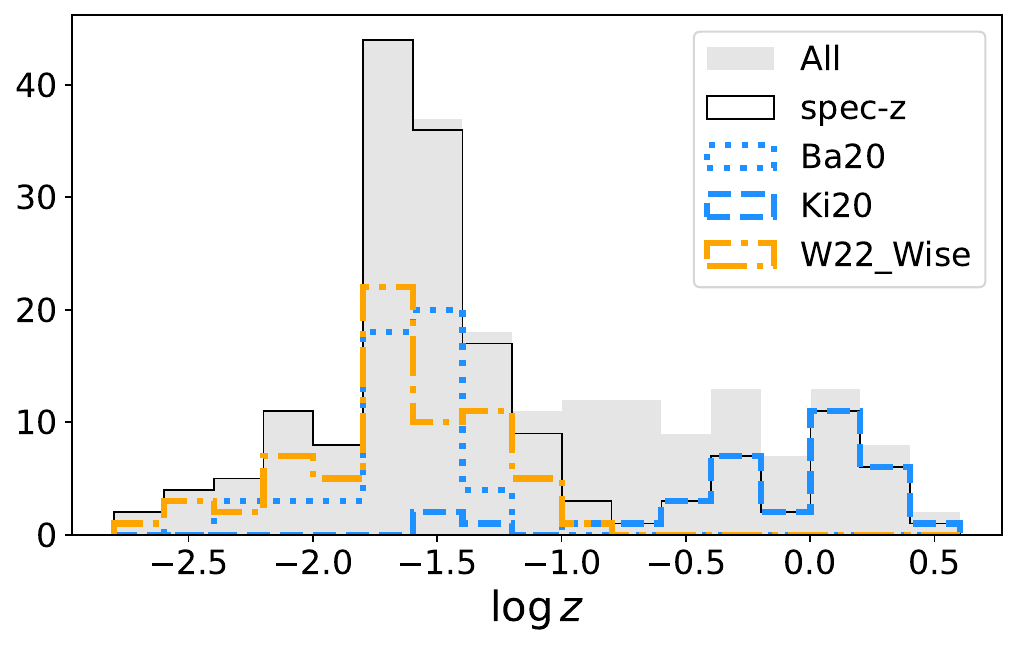}
    \includegraphics[width=0.83\columnwidth]{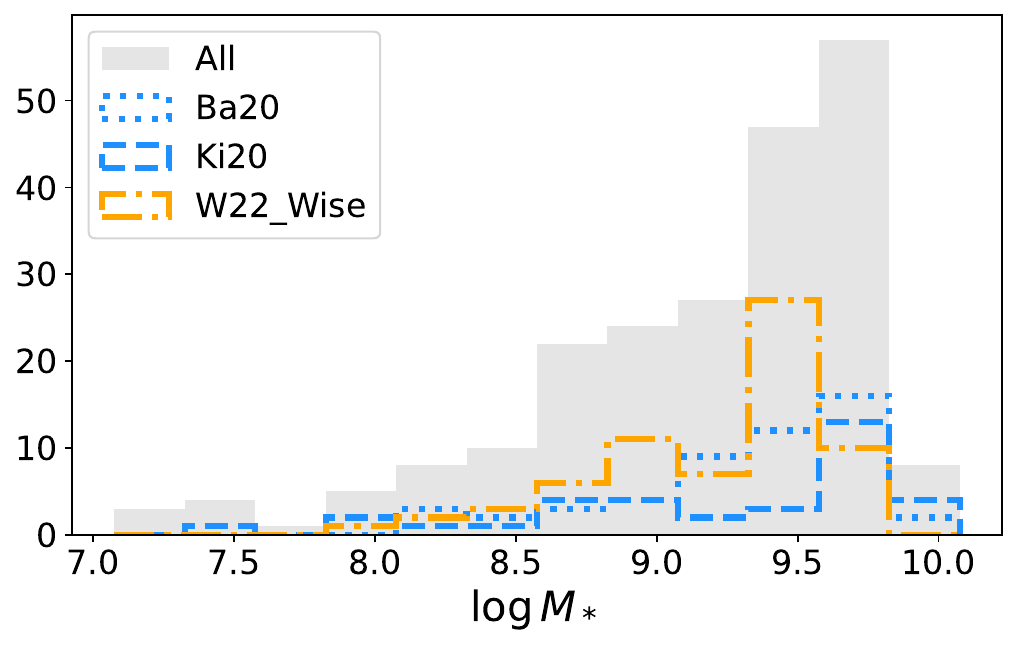}
    \includegraphics[width=0.84\columnwidth]{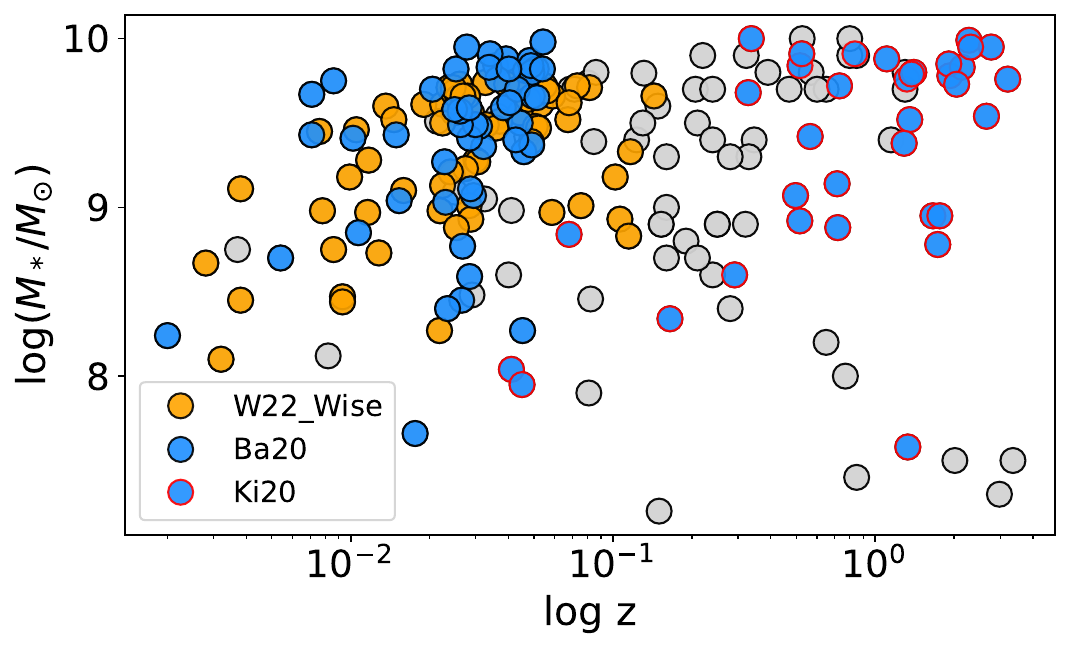}
    \caption{Top three panels: the r-band magnitude, redshift and stellar mass distributions of the parent sample adopted in this work (gray-filled histograms). We highlight the shallow low-z sample from \citet{Baldassare+2020:PTF} and the deep high-z sample from \citet{Kimura+2020:HSC} with blue dotted and dashed lines, respectively, to highlight the bimodality in magnitude and redshift within our sample. We also highlight with an orange dot-dashed line the bulk of the IR-selected MBHs \citep{Ward+2022:ZTF_Wise}. Bottom panel: the redshift versus stellar mass distribution.}
    \label{fig:hist}
    \end{figure}
    
    We draw samples of variable low-mass galaxies from the literature of optical \citep{Baldassare+2018:SDSS,Baldassare+2020:PTF,Kimura+2020:HSC,Ward+2022:ZTF_Wise,Burke+2022:DES,Shin+2022:intranight}, UV \citep{Wasleske+2022:Galex} and IR \citep{Secrest+2020:Wise,Ward+2022:ZTF_Wise,Harish+2023:Wise} studies. Albeit using slightly different methods and with different datasets and observing bands, all these works have performed similar searches for significant stochastic variability from the nuclei of dwarf galaxies, indicative of the presence of a MBH in their nuclei. The observed photometric light curves obtained from difference imaging are usually tested against a damped random walk model for AGN-like accretion variability \citep[e.g.,][]{Kelly+2009:DRW,Butler+2011:sigma}. As the emission from the galaxy is subtracted out, this technique has proved effective in finding faint nuclear AGN in dwarf galaxies, which would be otherwise missed with optical spectroscopy searches \citep{Baldassare+2018:SDSS,Baldassare+2020:PTF}, likely because these MBHs are not accreting close to the Eddington limit. However, the low-level variability does indicate that some level of accretion is happening in these nucleu, which implies that these MBHs are expected to emit X-rays. This makes the perfect sample for testing the synergies with UVOIR photometric surveys and eROSITA. The inhomogeneous and incomplete nature of the resulting galaxy sample is not concerning for the goal of this work, which is to compile a collection of dwarf galaxies with independent evidence of black hole activity in order to calibrate X-ray results in an informed way. Therefore, we assume that in this sample of variability-selected MBH candidates, the occupation fraction, namely the fraction of galaxies with a MBH seed in their center, and active fraction, namely that of galaxies with an active \citep[i.e. accreting, e.g.][]{Pacucci+2021:MBHs} black hole, are both one.
    
    The only selection criterion we perform on these datasets is a cut on stellar mass at $10^{7}\leq M_{*}\leq10^{10} M_{\astrosun}$ to select low-mass galaxies, taking $M_*$ from the above-mentioned literature or their parent samples. If information on the goodness of fit that yielded $M_*$ was found, it was used to filter $M_*$ by fit quality. For instance, we selected galaxies from \citet{Kimura+2020:HSC} with a reduced $\chi^2<10$ from SED fitting at all redshifts and additionally imposing a cut at $\chi^2<5$ at redshifts $z>1$, using the goodness of fit reported in \citet{Leigle+2016:cosmosMstar}. A more stringent criterion is used at higher redshift, where at fainter magnitudes (hence stellar masses) the same reduced $\chi^2$ can be obtained with a lower number of available filters. From \citet{Burke+2022:DES} we made use of $\Delta  \chi^2 $, which refers to the difference between the goodness of fit using the AGN template alone and the AGN+galaxy SED fit. We selected low-$M_{*}$ galaxies i) with $\Delta  \chi^2 >2$ from their SED fitting and with any variability timescale, or ii) sources with rapid variability (characteristic timescale lower than 1.5 days, \citealp{Burke+2022:DES}) and with any $\Delta \chi^2$ (Table~3 of \citealp{Burke+2022:DES}; C. Burke, priv. comm.). No explicit selection in redshift and narrow- and broad-lines classifications was performed. Redshifts are adopted from the references in Sect.~\ref{sec:sample} and consist, to the best of our knowledge, of spectroscopic redshifts for the vast majority\footnote{Galaxies with photometric redshifts from \citet{Burke+2022:DES} are knowingly included. Only sources with spec-z will be used in the X-ray stacking analysis.}. 
    Estimates of black hole masses in these galaxies are often absent or very uncertain and typical scaling relations with $M_{*}$ are not well calibrated in this mass regime \citep{Reines+2015:coev}. Therefore we do not make any preselection on $M_{BH}$ and for the scope of this paper we generically refer to these galaxies as MBHs or MBH candidates. 
    

    \begin{figure*}[t]
    	\centering
    	\includegraphics[width=1.23\columnwidth]{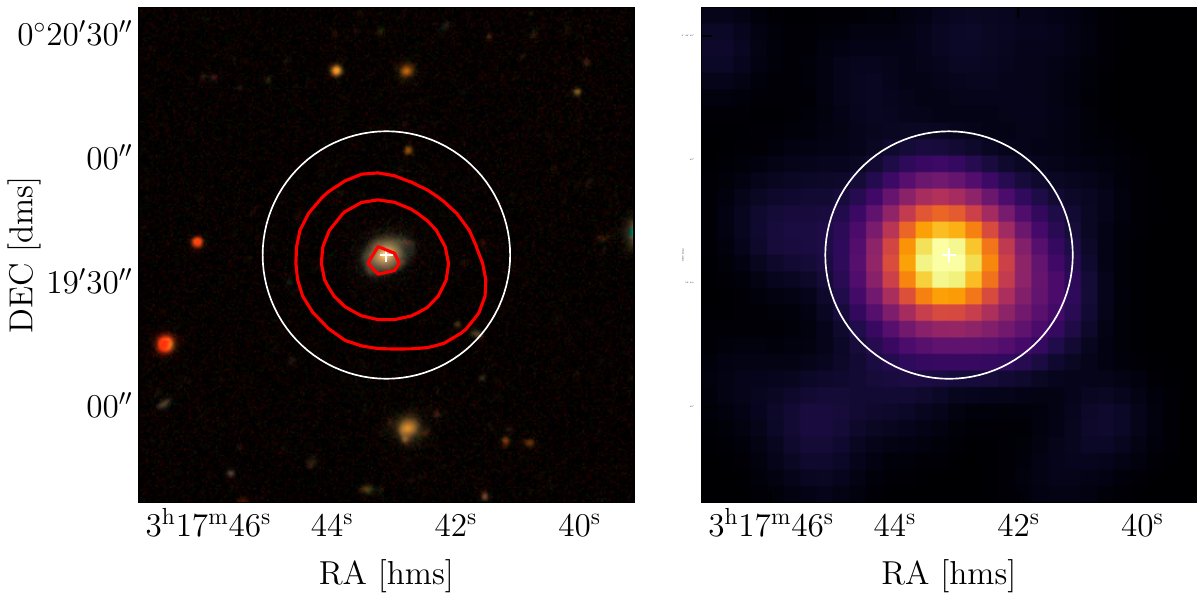}
    	\includegraphics[width=0.64\columnwidth]{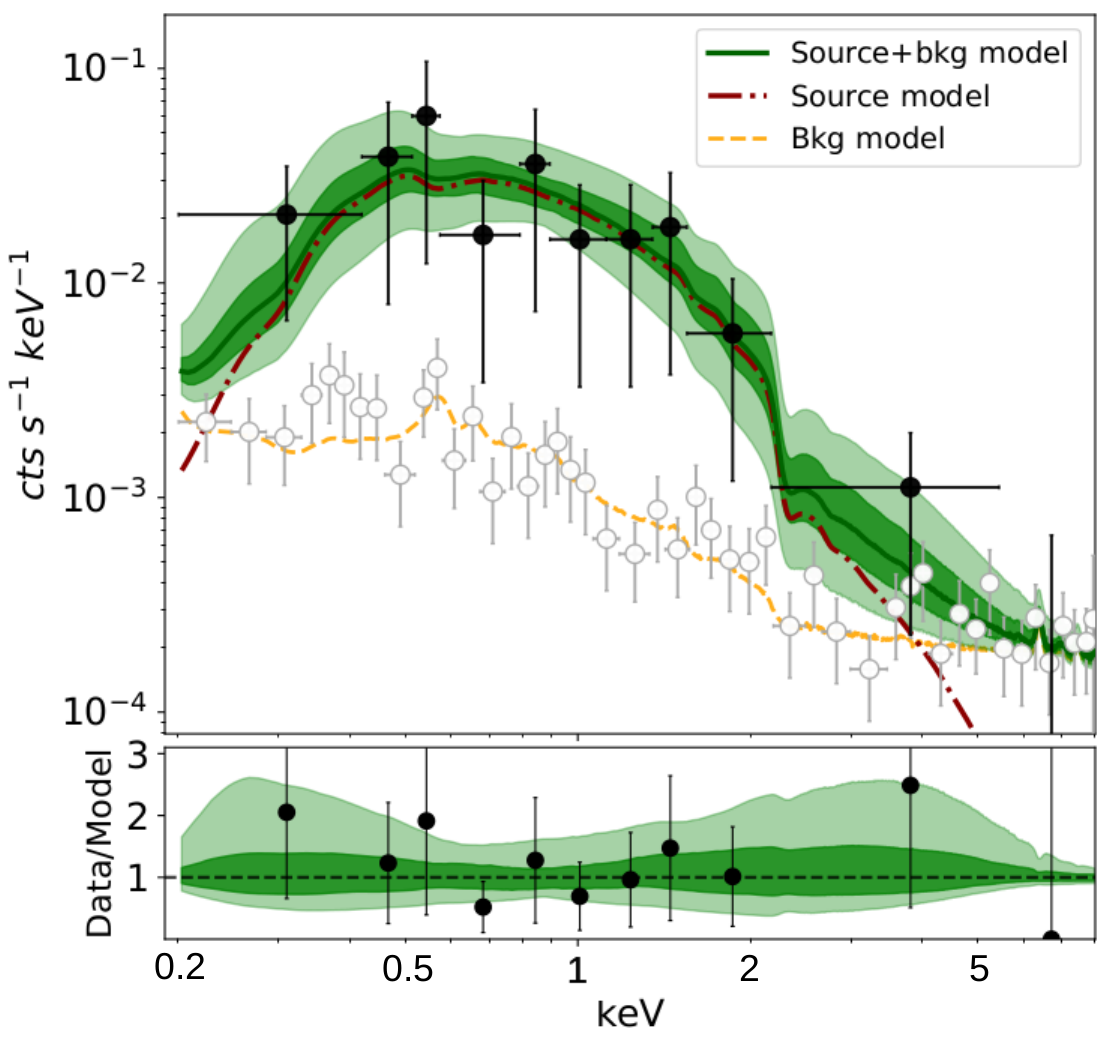}
    	\caption{Example of X-ray-detected MBH candidate in SDSS J031743.12+001936.8 at RA, Dec = (49.4296, 0.3269) and z=0.069, taken from \citep{Baldassare+2018:SDSS}. \emph{Left}: cutout of the DESI Legacy Imaging Surveys Data Release 10 [Legacy Surveys / D. Lang (Perimeter Institute)], centered at the input position. The white circle highlights the aperture of 30" used for X-ray products. Contours of the X-ray source are overlayed in red. \emph{Center}: eRASS:4 image centered at the input optical coordinates. Size and aperture circle correspond to those in the left panel. The positional accuracy of the X-ray centroid is 2", from the \texttt{POS\_COR} quantity \citep{Merloni+2023:erass} of the eRASS:4 catalog. \emph{Right}: X-ray spectrum of the X-ray source. Black points are source plus background data, empty gray points show the background alone. The power-law continuum model is shown by the dot-dashed red line, while the green line and related light green (dark green) shaded regions are the source plus background model median and 16th-84th (1st-99th) percentiles, respectively. The orange dashed lines shows the background model alone. In the lower panel, the data-model ratio is shown, following the format of the upper panel.}
    	\label{fig:examples}
    \end{figure*}

    A further obvious cut is the selection of galaxies in the German eROSITA hemisphere (i.e. Galactic latitudes between 179.944 and 359.944). 
    The total number of galaxies with stochastic nuclear variability in the German eROSITA footprint is 216. In particular for optically selected objects, we select three from \citet{Baldassare+2018:SDSS}, 52 from \citet{Baldassare+2020:PTF}, 35 from \citet{Kimura+2020:HSC}, six from \citet{Ward+2022:ZTF_Wise}\footnote{One duplicate in common between \citet{Ward+2022:ZTF_Wise} and \citet{Baldassare+2020:PTF} was removed.}, 46 from \citet{Burke+2022:DES}, and three from \citet{Shin+2022:intranight}. Then, 1 from \citet{Secrest+2020:Wise}, 66 from \citet{Ward+2022:ZTF_Wise}, 1 from \citet{Harish+2023:Wise} for infrared-selected MBHs and 3 from \citet{Wasleske+2022:Galex} for UV-selected ones. The total is thus 145 from optical photometry searches, 68 from the infrared and 3 from the UV. We show the r-band magnitude, redshift and stellar mass distribution of the entire parent sample in Fig.~\ref{fig:hist} in gray. The r-band magnitude and redshift distributions appear clearly bimodal. This is due to the presence of a large number of optically selected MBHs from \citet{Kimura+2020:HSC}, mostly high-z, and \citet{Baldassare+2020:PTF}, mostly low-z, with blue dashed and dotted lines, respectively. We highlight with an orange dot-dashed line the IR-selected MBHs from \citet{Ward+2022:ZTF_Wise}, to show that the bimodality in our sample is not due to the different wavebands. The different subsamples show marginal differences in the stellar-mass distribution instead (bottom two panels of Fig.~\ref{fig:hist}). The r-band magnitudes are selected from the SDSS NASA-Sloan Atlas sample\footnote{\href{https://www.sdss4.org/dr13/manga/manga-target-selection/nsa/}{Link to NSA catalog}} version 1.0.1 for the low-z subsample, whilst from the COSMOS Subaru/SuprimeCam \citep{Leigle+2016:cosmosMstar} for the high-z subsample.

    \section{X-ray analysis of eROSITA data}
    \label{sec:xray}
        
    Our method consists of systematically extracting X-ray photometry at the input UVOIR coordinates from the all-sky image of the first eROSITA survey (eRASS1) as well as from the cumulative image of the first four (eRASS:4). The former provides a show case for the data level being released \citep{Merloni+2023:erass}, while the latter for the deepest data level available full-sky to the German eROSITA Consortium. Images were extracted with the \texttt{evtool} task of the eROSITA Science Analysis Software System \citep[eSASS,][]{Brunner+2022:eSASS} from event files version 020. The algorithm to extract photometry makes use of the \texttt{Photutils} astropy package version 1.4.0 \citep{Bradley+2022:photutils1.4.0}. Photometry was extracted between $0.2-2.0\,$keV. We adopted a custom circular aperture of 30", corresponding to $\sim 75\%$ of the encircled energy fraction of eROSITA's point spread function in the adopted energy band. This source aperture is defined regardless on the presence of a detected X-ray source within. Background information is extracted from an annulus with inner and outer radii of 120" and 360", respectively. Every contaminating X-ray source in the field is masked out from both background and source apertures, although in the latter case only if the centroid of the X-ray contaminant is outside the source aperture. Potential contamination from within the source aperture, for instance due to ultra-luminous X-ray sources (ULXs), is studied a posteriori and discussed in Sect.~\ref{sec:discussion_det}. The coordinates of the masks are taken from the headers of eROSITA X-ray products extracted by eSASS. For a very small number of galaxies, the source aperture of 30" was masked out (entirely or $>70\%$) by a nearby bright or extended X-ray source. For eRASS1 images this is the case for 2/216 galaxies, while 8/216 for eRASS:4. This is due to the fact that eRASS:4 is deeper, therefore it contains more detected X-ray sources. We removed these from the parent sample (Sect.~\ref{sec:sample}) when computing detection fractions, thus the total number of galaxies with X-ray products is 214 for eRASS1 and 208 for eRASS:4.
    
	X-ray photometry yields counts in both the source and background apertures. From these, we compute the binomial no-source probability \citep[e.g.,][]{Luo+2017:Pb}, which yields the probability that the observed counts in the source aperture area are due to background fluctuations:
	\begin{equation}
	\label{eq:Pb}
	P_B (X \geq C_S) = \sum_{X = C_S}^{C_T} \frac{C_T!}{X!\,(C_T - X)!}\,A^X\,(1-A)^{C_T-X}
	\end{equation}
	where $C_S$ are the counts in the source aperture, $C_T = C_S + C_B$ and $C_B$ are the counts in the background area. Whereas $A = 1 / (1 + A_B/A_S)$, with $A_B$ and $A_S$ being the area of background and source apertures, respectively. We note that this area includes masks, therefore it is not always the full circle or the full background annulus regions as defined in input. $P_B$ can be calibrated in absolute sense only with simulations. For this, we use the ``digital twin'' of eRASS1 from \citet{Seppi+2022:erass1_simul}, which contains realistic populations of clusters and AGN. \citet{Seppi+2022:erass1_simul} ran source detection with the eSASS on the simulated sky, including the aperture photometry task \texttt{APETOOL}\footnote{\href{https://erosita.mpe.mpg.de/edr/DataAnalysis/apetool_doc.html}{Link to \texttt{APETOOL}}} \citep{Brunner+2022:eSASS}. From the simulations we know real and spurious sources that the detection algorithm finds and from \texttt{APETOOL} we know their counts\footnote{The impact of using a slightly different algorithm for aperture photometry is assumed to be negligible.} hence $P_B$. Here, we adopt as threshold for a significant detection $P_B = 0.0003$, which corresponds to $1\%$ of spurious fraction in the eRASS1 simulation. As a sanity check, we numerically computed on a one-dimensional grid in count rate the Poisson probability mass function (PPMF) from the detected counts using the \texttt{scipy} Python package \citep{Virtanen+2020:SciPy}. We compute count rate PPMFs for the source contribution alone, background alone and both source plus background. The PPMF for total (source plus background) and background-only count rates are compared and a detection is obtained when the two distributions are not compatible within 3$\sigma$, using the 1st and 99th percentiles of the related distributions. We verified that the two methods give the same number of significant detections. We note that we adopt $P_B <= 0.0003$ for detections in eRASS:4 as well, despite the value being calibrated for eRASS1. We expect minor differences for the purposes of this work, as the $P_B$ and PPMF detection criteria match for eRASS:4 as well.
	
	Spectra and light curves were extracted from the masked X-ray apertures of all sources, detected or undetected, using the \texttt{srctool} task in eSASS \citep{Brunner+2022:eSASS}. Spectral analysis is performed with the Bayesian X-ray Analysis software (BXA) version 4.0.5 \citep{Buchner+2014:BXA}, which connects the nested sampling algorithm UltraNest \citep{Buchner2019:mlf, Buchner2021:ultranest} with the fitting environment XSPEC version 12.12.0 \citep{Arnaud+1996:xspec}, in its Python version PyXspec\footnote{\href{https://heasarc.gsfc.nasa.gov/docs/xanadu/xspec/python/html/index.html}{Link to PyXspec}}. We adopted two simple continuum models, both with absorption fixed at the Galactic column density from HI4PI \citep{HI4PI+2016:HI4PI} and redshifted to rest-frame using the available redshifts: an accretion disk model, \texttt{zashift(diskbb}), and a power-law, \texttt{zpowerlw}. For the rest of this work, we adopt the \texttt{zpowerlw} model to quote flux and luminosity. For the detected sources, it is in the vast majority the model with higher Bayesian evidence from the BXA fit and data-model ratio residuals were visually confirmed to be acceptable. The choice has a negligible impact, also for the upper limits of the non detected sources. Flux and luminosity are computed in the rest-frame $0.2-2.0\,$keV band. We quote median and 1st and 99th percentiles ($\sim3\sigma$) from fit posteriors, unless otherwise stated, for fit parameters, flux and luminosity. For non-detections ($P_B$>0.0003), as defined above, we quote upper limits using the 99th percentiles of the fit posteriors, unless otherwise stated.

    \begin{figure*}[t]
    	\centering
    	\includegraphics[width=0.85\columnwidth]{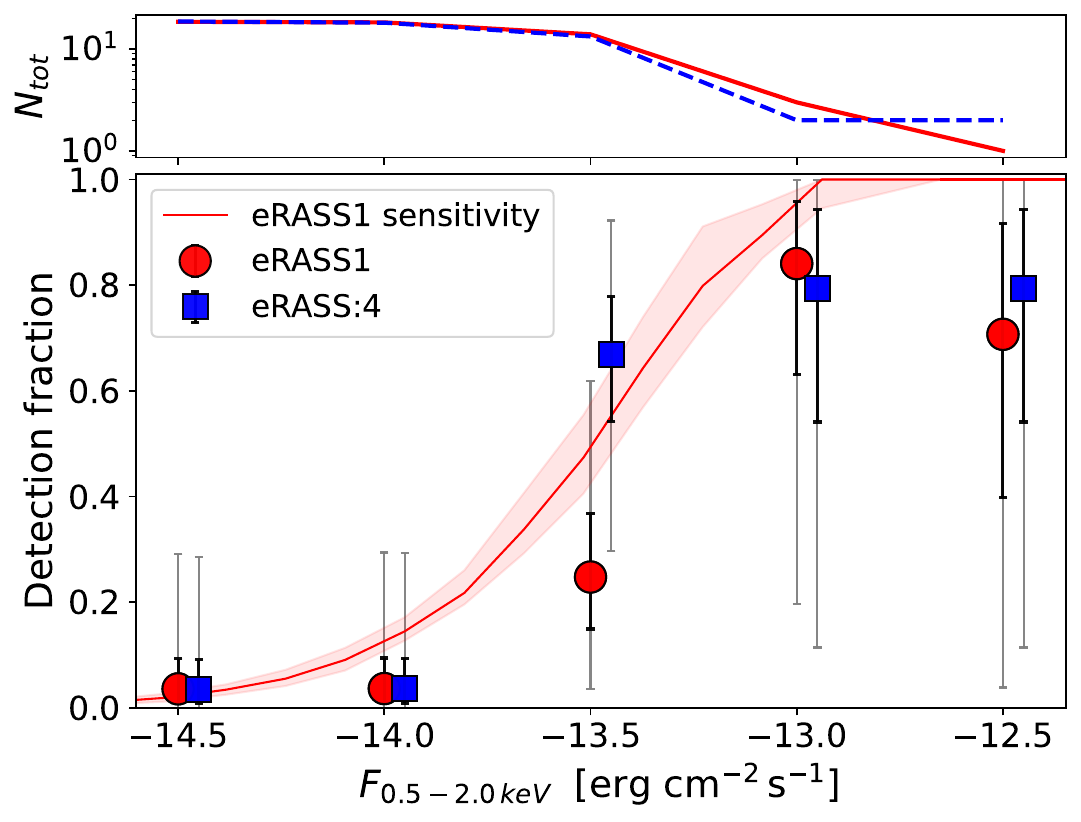}   
		\includegraphics[width=0.85\columnwidth]{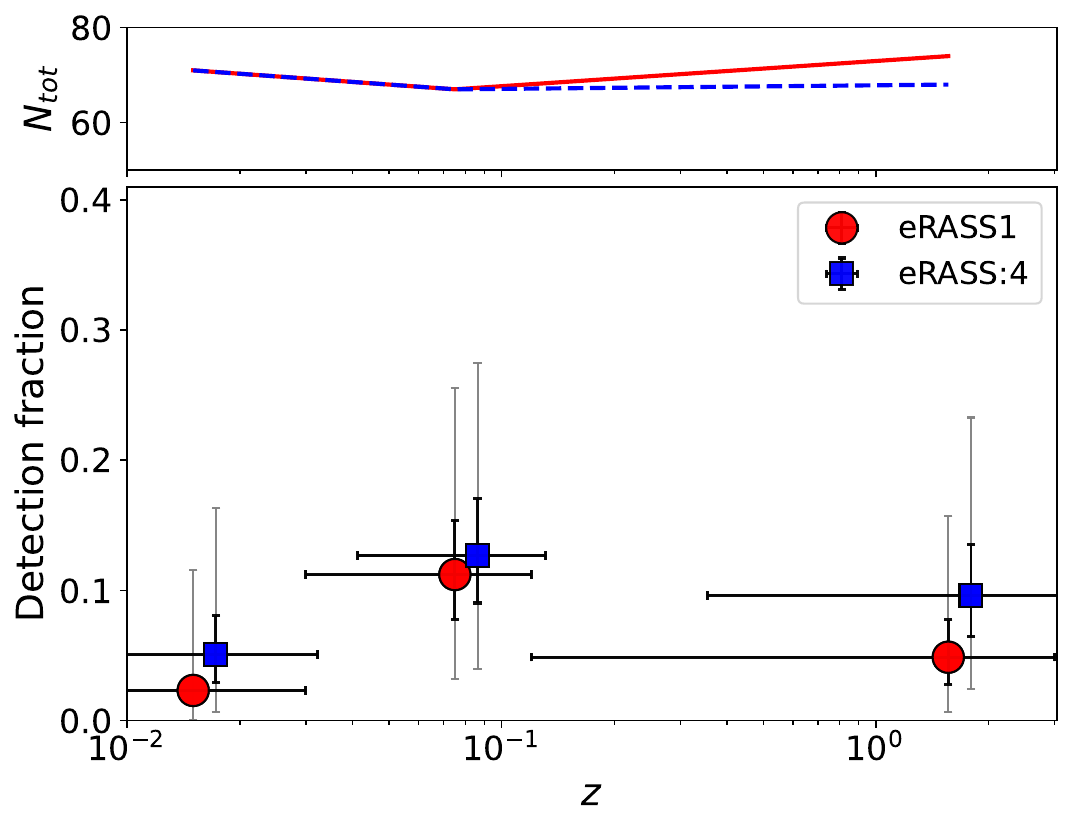}
		\includegraphics[width=0.85\columnwidth]{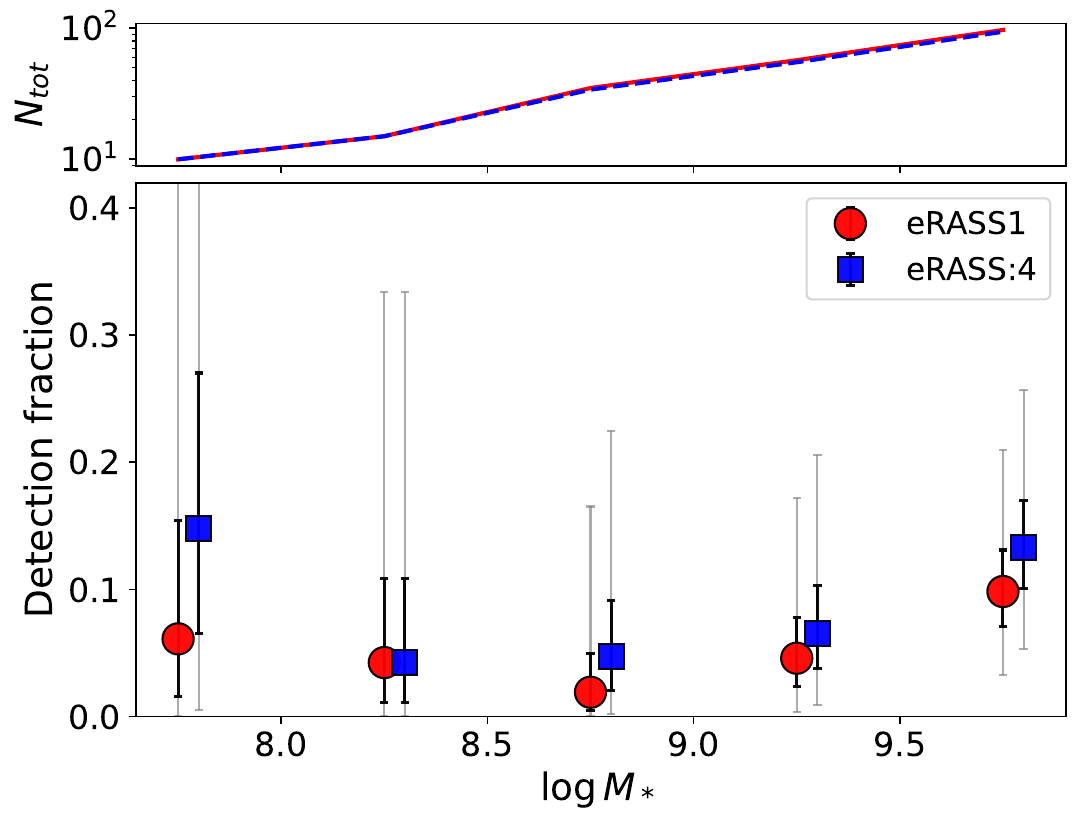}
		\includegraphics[width=0.85\columnwidth]{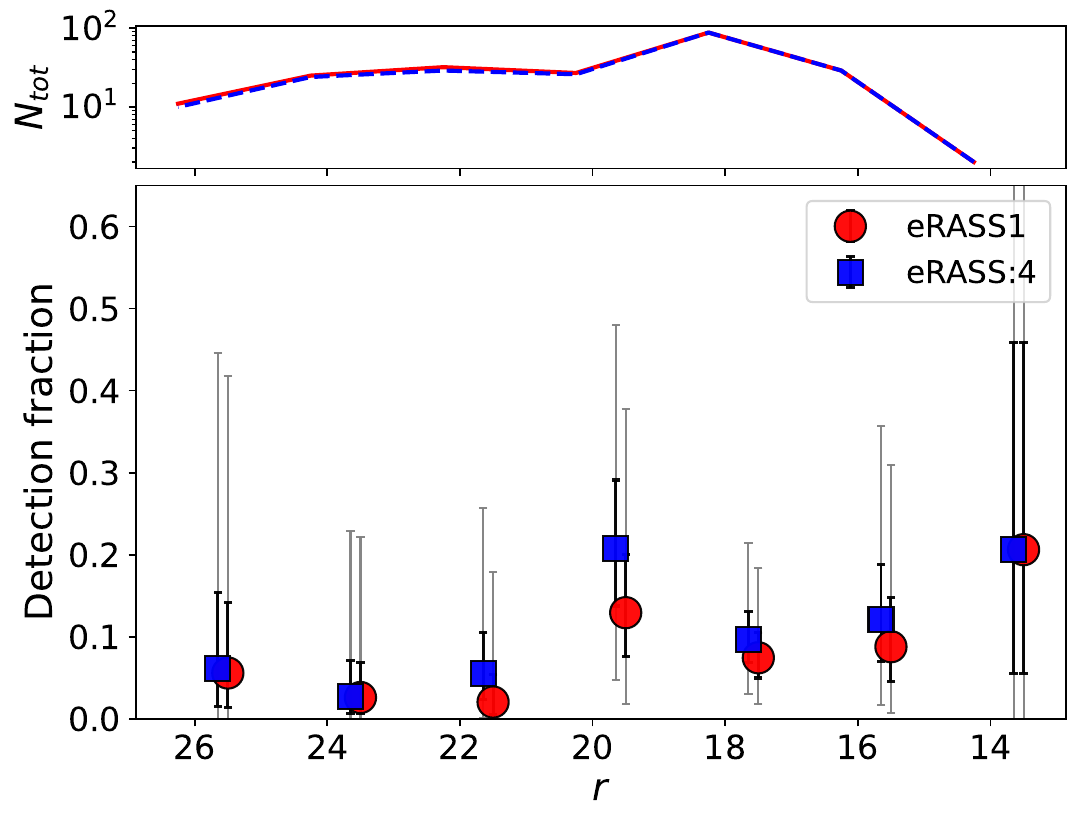}
    	\caption{Fraction of input galaxies detected in eRASS1 (red) and eRASS:4 (blue) as a function of X-ray flux (\emph{top left}), redshift (\emph{top right}), stellar mass (\emph{bottom left}) and r-band magnitude (\emph{bottom right}). Different symbols, between eRASS1 and eRASS:4, are slightly shifted horizontally for illustration purposes. In the top left panel, the red dotted line shows the eRASS1 sensitivity curve \citep{Seppi+2022:erass1_simul}. In all subplots, upper subpanels show the number of galaxies in each bin. $1\sigma$ ($3\sigma$) binomial confidence intervals \citep{Cameron+2011:beta} are shown in black (gray).}
    	\label{fig:detfrac}
    \end{figure*}
    
    Finally, we performed stacking analysis of non-detections following the method presented in \citet{Comparat+2022:stacking}. Here, we outline the main steps. For each galaxy, the physical distances between X-ray photons and the galaxy ($R_{\rm kpc}$) are calculated according to the spectroscopic redshift of the galaxy and observed angular distance. We retrieve photons within $0.5-2.0\,$keV and within 50\,kpc of each galaxy and create a photon cube saving the positions, the distance to the associated galaxy (angular and physical, $R_{\rm rad}, R_{\rm kpc}$), the exposure time $t_{\rm exp}$, the observed energy $E_{\rm obs}$, the emitted energy $E_{\rm rest}=E_{\rm obs}*(1+z)$, and the effective area $A_{\rm eff}$. These photons within 50\,kpc will be used for both source and background estimates, as detailed below. All the X-ray-detected sources in the field are masked out and the related correction factor of the area ($A_{\rm corr}$) is calculated as a function of $R_{\rm rad}$ or $R_{\rm kpc}$. 
    We then merge the photons around the galaxies of interest and calculate the surface brightness ($I_{X}$) of the stacked image:
    \begin{equation}
        I_{X}=\Sigma\frac{A_{\rm corr}4\pi D_{\rm g}^2 E_{\rm rest}} {A_{\rm eff}t_{\rm exp}}\frac{1}{N_{\rm g}},
    \end{equation}
    where $D_{\rm g}$ is the luminosity distance of the galaxy and $N_{\rm g}$ is the number of stacked galaxies. This profile is then integrated up to a given distance (angular or physical) to yield a median X-ray luminosity of the stacked image, with related Poisson statistical uncertainty. \citet{Comparat+2022:stacking} estimated that the uncertainty due to the source-masking in the stacking procedure amounts to at most a $\sim2\%$ uncertainty on the number of events. To be conservative, we apply a $2\%$ systematic uncertainty to the measurements. We integrate up to 10\,kpc unless otherwise stated. This scale is a few times larger than the typical effective radius, or half-light radius, of galaxies below $\log M_* = 10$ \citep[e.g.,][]{Gadotti+2009:reff}, therefore the relevant scale is the much larger eROSITA's PSF. An integration up to 10kpc ensures that the eROSITA PSF is contained fully within the integration bounds for sources at the median redshift of the $z<0.1$ subsample, whilst minimizing the presence of possible stacked signal from the outskirts of galaxies. Furthermore, we check that the stacked image detection or non-detection remains such changing the integration distance, and by visualizing the profiles to exclude that the detection is not driven solely by spurious signal in a single off-centered annulus. The background is calculated taking the median value of the signal between $15<R_{\rm kpc}<50$ and it is subtracted from each annulus during integration. We visualize that the stacked signal between $15<R_{\rm kpc}<50$ is constant.  We conservatively check that a detection remains such also if the 84th percentile of the signal within $15<R_{\rm kpc}<50$ is used as background estimate and if the lower integration bound is moved inward or outward from 15\,kpc. If the stacked signal is compatible, within its uncertainties, with the background estimate, we quote the background-subtracted upper value of the luminosity integral as upper limit. An example is provided in Fig.~\ref{fig:stack}, where only the signal shown in red represents a detection, whilst that in green is compatible with background.
    
    \section{Results}
    \label{sec:results}

    \subsection{Detection fraction}

    
    We obtain that $5.1_{-1.1}^{+2.0}\%$ (11/214) of the dwarf galaxies are detected in eRASS1 and $8.2_{-1.5}^{+2.3}\%$ (17/208) in the deeper eRASS:4 (see Sect.~\ref{sec:xray}). The median fraction and 1$\sigma$ binomial confidence intervals are inferred from the related quantiles of the beta distribution from \citet{Cameron+2011:beta}. In particular, we detected in eRASS1 (eRASS:4) 3 (3) galaxies from \citet{Baldassare+2018:SDSS}, 3 (4) from \citet{Baldassare+2020:PTF}, 1 (4) from \citet{Burke+2022:DES}, 0 (1) from \citet{Shin+2022:intranight} and 4 (5) from the WISE-selected sources in \citet{Ward+2022:ZTF_Wise}. In eRASS:4, detection fractions of $9.2_{-1.9}^{+3.2}\%$ and $7.2_{-2.0}^{+4.4}\%$ are obtained for the optically- and IR-selected galaxies, respectively, thus they are compatible within uncertainties. We show an example of a detected source in Fig.~\ref{fig:examples} to showcase our methodology. The input coordinates and the adopted aperture are shown with a white circle in both left and central panels, showing the optical and X-ray cutouts, respectively. The right panel shows the source plus background spectrum and related model lines and contours. We report $P_B$ and X-ray luminosity ($L_{0.2-2.0\,keV}$) for all detected and undetected dwarf galaxies, for both eRASS1 and eRASS:4, in Table~\ref{tab:results}. For a consistency check, we compared our eRASS1 results with the official eRASS1 catalog released in \citet{Merloni+2023:erass}, matching the optical coordinates in input within 30", the circular aperture used here for X-ray products. All the 11 eRASS1 sources found in this work are present in the official catalog with compatible fluxes. Four sources which are considered undetected in this work are present in the official eRASS1 catalog (namely 1eRASS J130716.6+133904, 1eRASS J032845.6-271113, 1eRASS J003429.2-432056, 1eRASS J085125.9+393541; \citealp{Merloni+2023:erass}). They have $P_b$ spanning between 0.0004 and 0.003, therefore they are marginally below our $P_b=0.0003$ threshold. The three sources above $P_b>0.001$ have detection likelihoods between 6-8 in the official catalog (\texttt{DET\_LIKE\_0}; see \citealp{Merloni+2023:erass}), which corresponds to a false detection rate between $\sim4-14\%$ \citep{Seppi+2022:erass1_simul}. However, these four sources are all detected in the deeper eRASS:4 image with our method. Therefore, they are most likely real sources and this comparison simply implies that our algorithm and chosen $P_b$ threshold are on the conservative side. As a matter of fact, we adopted a threshold of $P_b=0.0003$ to ensure a lower spurious fraction of $\sim1\%$.
       
    We show the detection fraction as a function of X-ray flux (in the rest-frame $0.5-2.0\,$keV band) in the top left panel of Fig.~\ref{fig:detfrac}. Different symbols, between eRASS1 and eRASS:4, are slightly shifted horizontally for illustration purposes. In order to compute the evolution of detection fraction as a function of X-ray flux, we included non detected galaxies in the plot by extracting 100 random values from their unconstrained flux chains. In this way, each source may enter different bins at each iteration. We averaged over these 100 iterations, therefore uncertainties include the fact that non-detections are spread across more bins. As they would be more likely extracted in the lower flux bins, their binomial uncertainties are smaller than the high-flux bins (the average numbers per bin are shown in the upper subpanel). Non detected sources with a flux fainter than the lowest bin (-14.75, -14.25) are not present in any bin at a given iteration. The evolution of detection fraction as a function of X-ray flux can be compared with eROSITA's sensitivity. For eRASS1, we can use the simulations from \citet{Seppi+2022:erass1_simul} which provide the eRASS1 sensitivity curve. Since simulations were done for each sky tile, we can compute the eRASS1 sensitivity at the locations of all sources in our parent sample. We show the median (with related 16th and 84th percentile contours) of this distribution with a solid red line in the top left panel of Fig.~\ref{fig:detfrac}. We note that eRASS1 MBH detections from this work lie below the sensitivity curves from simulations at low and intermediate fluxes. 
    This might suggest that not all the UVOIR-variable MBHs in input are intrinsically above an X-ray flux of $\log (F_X/($erg\,s$^{-1}))\sim-14.5$, used in the plot at the lower end. 
    We note that, however, we do not expect all MBHs in the sample to be intrinsically above an X-ray flux of $\log (F_X/($erg\,s$^{-1}))\sim-14.5$, given that our sample includes also high-redshift sources (around half of the input sample is above $z\sim0.04$), for which such a threshold flux would correspond to a significant intrinsic luminosity. Indeed, it is unreasonable for all the MBHs above this redshift to be above an intrinsic luminosity of $\sim1.2\times10^{40} (D_L/D_{L_{0.04}})^2\,$erg\,s$^{-1}$, for a luminosity distance $D_{L_z}$. The situation marginally improves when filtering the top left panel of Fig.~\ref{fig:detfrac} below $z\sim0.04$, although only due to the even larger error bars which is merely due to the decrease of sample size. Further, the top right panel of Fig.~\ref{fig:detfrac} shows the observed detection fraction as a function of redshift in three bins with roughly equal number of galaxies. We note no singificant difference across the bins. We conclude that the incompatibility between observations and simulations is likely not uniquely a redshift effect and it will be investigated and discussed further in Sect.~\ref{sec:discussion_xweak}.

    \begin{figure}[t]
	\includegraphics[width=0.99\columnwidth]{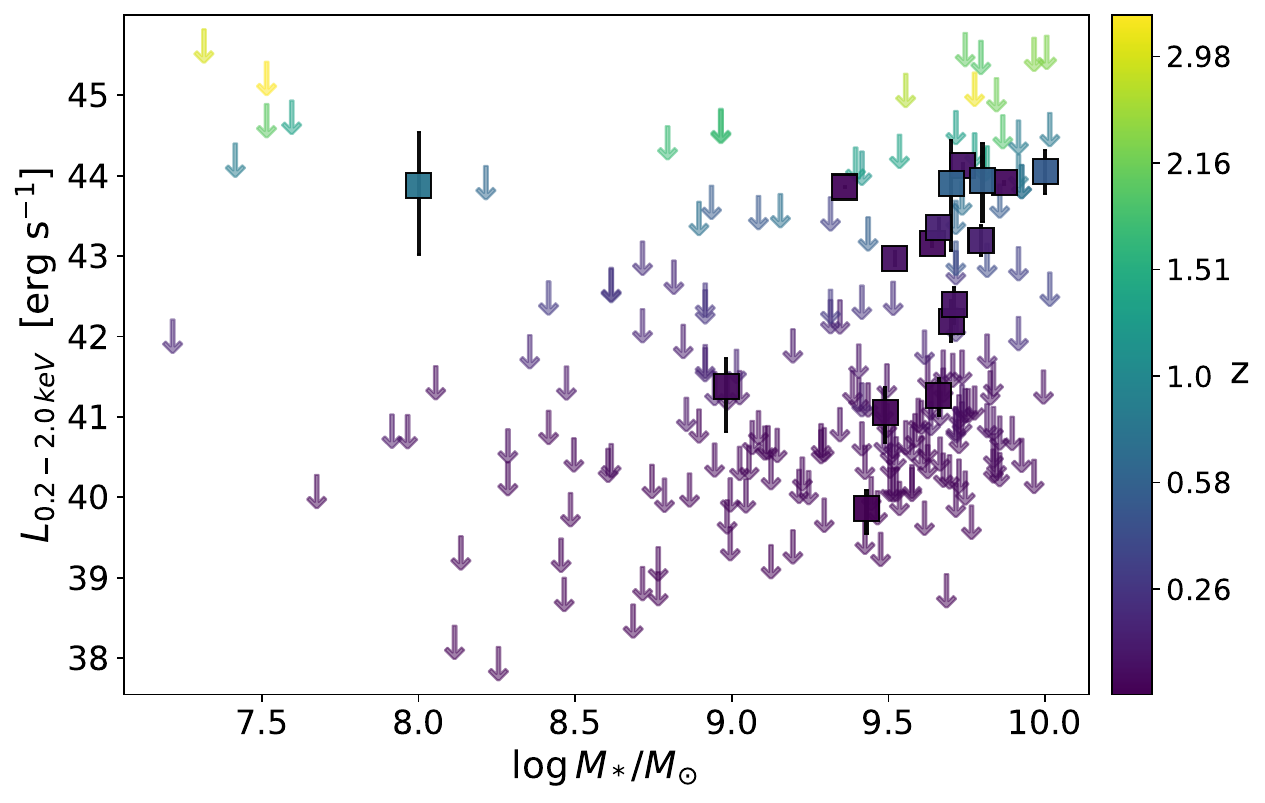}
        \\ \quad \\
        \includegraphics[width=0.99\columnwidth]{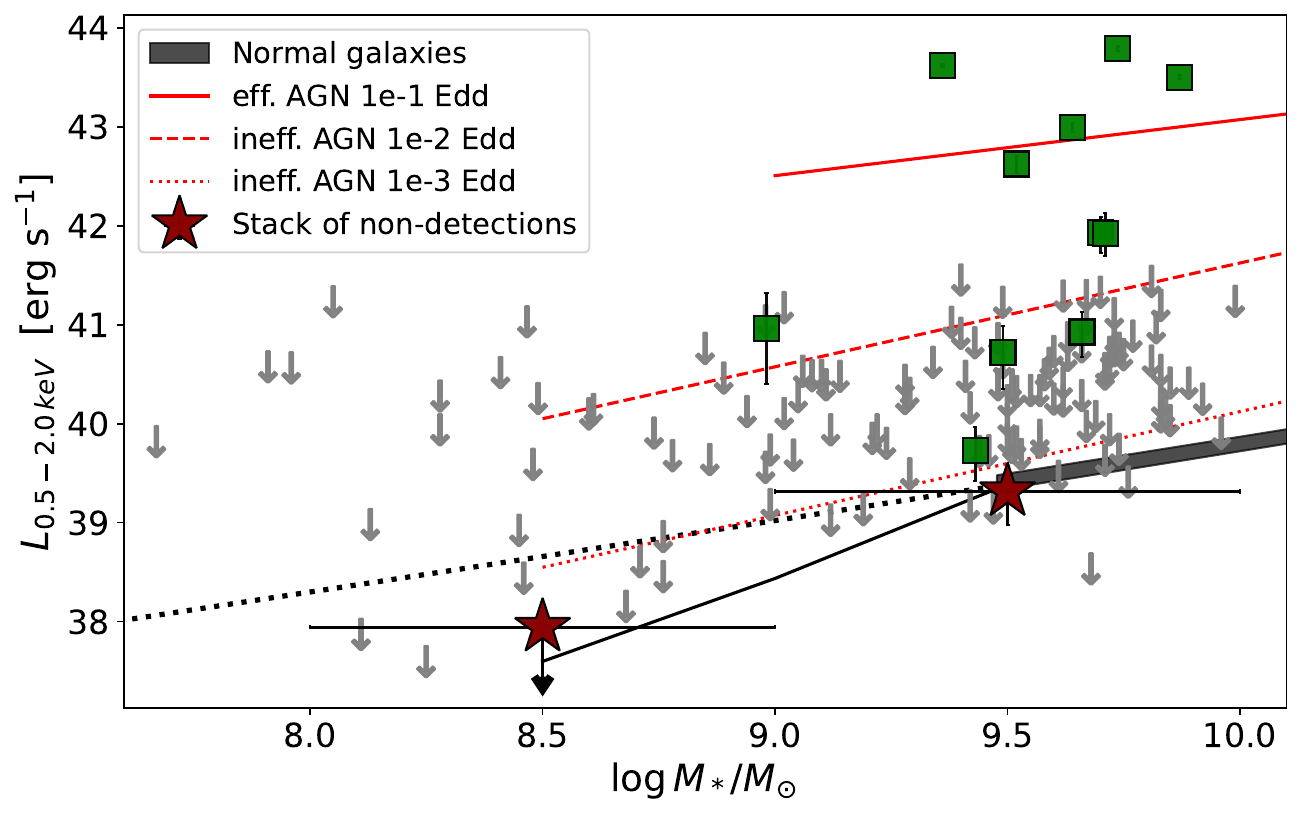}
	\caption{X-ray luminosity from eRASS:4 as a function of host galaxy stellar mass. In both panels, detected MBHs are shown with squares, whilst non-detected ones with arrows. In the top panel, all the sources are shown and are color-coded as a function of logarithmic redshift. In the bottom panel, only sources with $z<0.1$ are shown: non detections (gray arrows) were stacked in two mass bins (highlighted by the x-axis error bars) and the related X-ray luminosity estimates are shown with red stars. We show the soft X-ray luminosity predicted for normal galaxies (black shaded contour and solid line) and AGN at different accretion states in the same stellar mass range (see Sec.~\ref{sec:mstar} for details).  In particular, predictions for radiatively-efficient AGN are shown with a solid red line, while predictions for AGN accreting at 1e-3 (1e-4) of $L/L_{edd}$ are shown with a dashed (dotted) line.}
    \label{fig:lxmstar}
    \end{figure}

    \subsection{Trends with the galaxy's stellar mass}
    \label{sec:mstar}
    
    From the bottom left panel of Fig.~\ref{fig:detfrac} we note a slight increase of detections with increasing stellar mass, although all values are compatible within $3\sigma$ uncertainties. In both eRASS1 and eRASS:4, the overall detection fraction of $\sim5\%$ and $8\%$, respectively, is compatible with those estimated in the single stellar mass bins, within uncertainties. 
    Based on this, we obtain that we can expect to detect from any future UVOIR variability survey, with similar characteristics to the ones considered here, a fraction on the order of $\approx5\%$ ($\approx8\%$) in eRASS1 (eRASS:4) at least above $\log M_*\sim8.5$. We show the eRASS:4 detections and non-detections in the luminosity-stellar mass plane (Fig.~\ref{fig:lxmstar}). The top panel shows the full sample, where the low-z and high-z populations (e.g., see the top-middle panel of Fig.~\ref{fig:hist}) are clearly separated. We note an outlier in the X-ray-detected source around stellar mass of $\sim10^8 M_{\odot}$. This estimate from \citet{Burke+2022:DES} comes with a high statistical uncertainty ($\sim0.5\,$dex) and the marginal increase in $\Delta \chi^2$, between the AGN template alone and the AGN+galaxy SED fit, implies large systematics which hinder a reliable interpretation of the stellar mass value (C. Burke, priv. comm.). 

    \begin{figure}[t]
	\includegraphics[width=0.9\columnwidth]{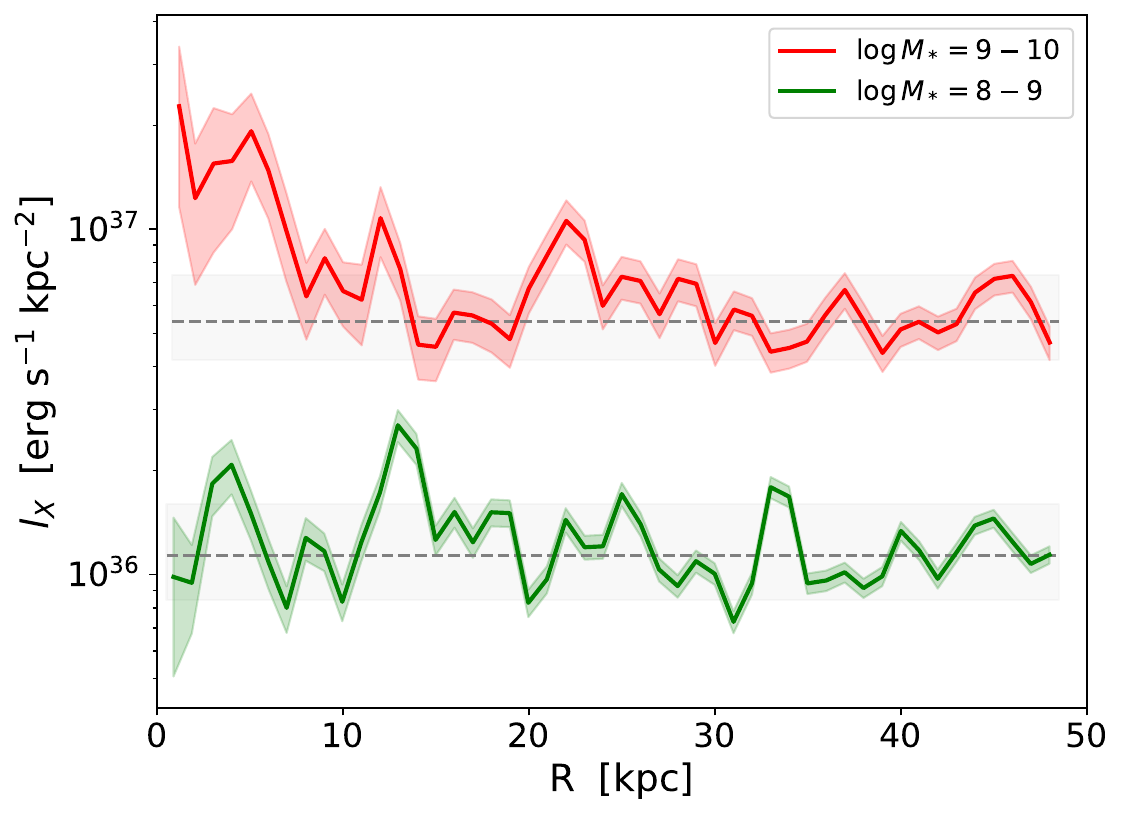}
	\caption{Emission profiles from the stacked images in two $M_*$ bins, $\log M_* = 8-9$ (green) and $\log M_* = 9-10$ (red). The source signal is integrated up to 10\,kpc, whilst the background is estimated from the median (or, the 84th percentile value, conservatively) of the emission between 15 and 50\,kpc (see Sect.~\ref{sec:xray}). The stack contains signal above background only in the $\log M_* = 9-10$ bin.}
    \label{fig:stack}
    \end{figure}
    
    In general, our sample is rather heterogeneous and obtained through different selection methods (Sect.~\ref{sec:sample}), therefore for further analysis and data-model comparisons in the $L_X-M_*$ plane we only use the subsample of 134 galaxies below $z<0.1$ (e.g., see the bottom panel of Fig.~\ref{fig:lxmstar}). This selection allows us to use an homogeneous low-z population and magnitude distribution (see Fig.~\ref{fig:hist}). In particular, we stacked the $0.5-2.0\,$keV eRASS:4 images of the 121 undetected sources below of $z<0.1$, using only spectroscopic redshifts. We stacked two sets of images in two $M_*$ bins, $\log M_* = 8-9$ and $9-10$, which contain 30 and 91 undetected galaxies respectively. The low mass bin stack is undetected, with an upper limit at $L_{0.5-2.0\,keV}<9\times 10^{37}\,$erg\,s$^{-1}$, whilst in the high-mass bin we obtain $L_{0.5-2.0\,keV} = (2.1\pm1.1)\times 10^{39}\,$erg\,s$^{-1}$. The profiles are shown in Fig.~\ref{fig:stack} and they are represented with dark red stars in the bottom panel of Fig.~\ref{fig:lxmstar}.

    With the aim of interpreting the observed X-ray luminosities, we compare them with predictions of both AGN and normal galaxies. We computed the predicted $0.5-2.0\,$keV X-ray luminosity from X-ray binaries in normal galaxies following \citet{Lehmer+2016:LXRB} and added the diffuse hot gas component due to the ISM, relevant in the soft X-rays, following \citep{Mineo+2012:hot}. We adopt the stellar mass from our parent sample and use the star formation main sequence \citep{Whitaker+2012:sfrms}, for simplicity, to obtain the star formation rate (SFR) for this plot. We note that for starburst galaxies, this would be an underestimation of SFR. This prediction is shown with the black thick line in the bottom panel of Fig.~\ref{fig:lxmstar}, with the thickness spanning the prediction for the minimum ($z=0$) and maximum ($z=0.1$) redshifts of the galaxies in the panel. Below $\log M_* \sim 9.5$ and below SFR\,$\sim 2 M_{\odot}$/yr the relation is known to be inaccurate, due to the fact that the galaxy prediction relies on fully-populated X-ray binaries luminosity functions \citep{Gilfanov+2004:xrbs,Lehmer+2019:xrb}, which would not apply in this regime. The black dotted line can be used as guide for the eye, in case this relation still holds on average \citep[e.g.,][]{Kouroumpatzakis+2020:gal}, albeit with significant scatter (e.g. Kyritsis et al., in prep). If stochastic sampling implies higher difficulty in observing luminous sources reducing the average luminosity per galaxy, the dotted line would be an overestimate. We approximate this by artificially decreasing the dependency on $M_*$ and SFR \citep[e.g. Fig. 16 in][]{Lehmer+2019:xrb}, thus the predicted X-ray luminosity, and we show this with a solid black line in Fig.~\ref{fig:lxmstar}. 
    
    Furthermore, we computed the predicted AGN soft X-ray luminosity as a function of galaxy stellar mass by interpolating scaling relations and spectral energy distributions (SEDs) common to more massive AGN. Since typical scaling relations are calibrated in the UV \citep[e.g.,][]{Arcodia2019:lxluv,Ruan+2019:aox}, but still hold for a wide range of optical frequencies \citep{Jin+2012:lxlo}, we adopt the bluest SDSS filter available, for simplicity. We obtained the observed u-band flux of our galaxies from the parent SDSS NASA-Sloan Atlas sample (\texttt{EL\_PETRO\_FLUX}). No K-correction was applied to these estimates, as they are intended as guide for the eye. We infer the AGN optical luminosity assuming accretion at $\sim0.1\times$, $\sim0.01\times$ and $\sim0.001\times L_{edd}$, assuming $M_{BH}=0.002 M_*$ and an optical bolometric correction of 0.1 \citep[e.g.,][]{Merloni2016:contrast}. Then we applied X-ray-to-optical scaling relations for radiatively-efficient \citep{Arcodia2019:lxluv} and -inefficient \citep{Ruan+2019:aox} AGN to infer the expected $2\,$keV luminosity, and finally converted to $L_{0.5-2.0\,keV}$ assuming a power-law spectrum with photon index 1.9. Quite interestingly, the detected MBHs (green squares) mostly align with the predictions of AGN accreting at $\approx0.01-0.1 L_{edd}$. However, we notice that the vast majority of the eRASS:4 $3\sigma$ upper limits lie well below these scaling relations. Most importantly, the X-ray luminosity estimates from their stacked images (dark red stars) are consistent with predictions of normal galaxies' non-AGN emission. We note that despite $M_*$ is a notoriously uncertain parameter, most upper limits would remain inconsistent with the AGN predictions even if they were biased low or high in stellar mass by as much as $\sim0.5-1.0$\,dex (e.g. along the x-axis of Fig.~\ref{fig:lxmstar}), and the stacks would likely be unaffected by a few erroneous stellar mass estimates. The nature of this X-ray weakness will be further explored in Sect.~\ref{sec:discussion_xweak}, by comparing X-rays to other wavebands as well.   
      
    \section{X-ray-detected dwarf galaxies}
    \label{sec:discussion_det}
    
    \subsection{Contaminants: the cumulative stellar-mass BHs population}
    
    We investigate the possible cumulative contribution to the X-ray-detected galaxies due to the stellar population, here intended as a contaminant, within the host galaxy of our MBH candidates \citep[e.g.,][for a recent review]{Gilfanov+2022:rev}. We use the term X-ray binary (XRB) for the collective contribution of both accreting neutron stars and stellar-mass black holes. Despite the difficulty of securely assessing contamination from XRBs for each galaxy, we can rely on well-known scaling relations that predict the expected X-ray luminosity from XRBs given the stellar mass and SFR in the galaxy. The mass of the stellar companion defines the classification between low- and high-mass XRBs. The former (latter) kind evolves slower (faster) and it is therefore traced by the total stellar content or $M_*$ (by recent star formation and SFR and both have to be taken into account \citep[e.g.,][]{Grimm+2003:xrbs}.    
    
    We compute the predicted X-ray luminosity ($L_{X,gal}$) in the $2-10\,$keV range from the cumulative XRB population in the host galaxy from  \citet[][their Eq. 15]{Lehmer+2016:LXRB}, which was calibrated in the Chandra Deep Field-South (CDF-S):
    \begin{equation}
    \label{eq:lehmer}
    L_{X,gal} = \alpha_0 (1+z)^{\gamma_0} M_* + \beta_0 (1+z)^{\delta_0} \text{SFR}
    \end{equation}
    with $(\log \alpha_0, \log \beta_0, \gamma_0, \delta_0) = (29.30, 39.40, 2.19, 1.02)$. For these calculations, we obtained individual SFR values from different sources: five galaxies match with the HECATE catalog \citep{Kovlakas+2021:HECATE} within 3", five with \citet{RamosPadilla+2022:SED}, one from \citet{Omand+2014:gal} and one from \citet{Chang+2015:sfr}; for the remaining sources SFR was obtained from UV \citep{Bianchi+2017:galex} and IR \citep{Cutri+2012:WISE} fluxes, following the prescription from \citet{Lehmer+2019:xrb}. These values span uniformly between $\sim1-100 M_{\odot}\,$yr$^{-1}$. For consistency with the SFR estimates, we used $M_*$ from these references for computing $L_{X,gal}$, if present, or the values in Table~\ref{tab:results} otherwise. Here, we neglect the contribution from hot diffuse gas due to the ISM to $L_{X,gal}$ since it is expected to be significantly lower than the faintest of our X-ray detections ($\sim7\times10^{39}\,$erg\,s$^{-1}$), even more so given the range of stellar masses in our sources and in the $\sim2-10\,$keV band. As a matter of fact, this contribution is $L_X/M_* \sim 10^{28}\,$erg\,s$^{-1}$\,M$_{\astrosun}^{-1}$ for early type galaxies \citep[e.g.,][]{Hou+2021:Lgas} and amounts to up to $\sim10\%$ of the observed luminosity for star-forming galaxies \citep[][Kyritsis et al., in prep]{Mineo+2012:hot}. Here, we ignore the known stochasticity of the galaxy prediction at low $M_*$ and SFR \citep{Gilfanov+2004:xrbs,Lehmer+2019:xrb}, for simplicity. The adopted scaling relations surely come with considerable uncertainties and intrinsic scatter, although one of the causes of this scatter at the bright end is the likely presence of X-rays from the MBH itself. A further source of contamination which we neglect here could be the cumulative emission from XRB from the nuclear star cluster (NSC), which is nearly ubiquitous in low-mass galaxies \citep{Neumayer2020:NSC,Hoyer+2021:nsc,hoyer+2023:nscs}. As standard scaling relations to estimate $L_{X,gal}$ try to exclude the point-like nuclear X-ray source, to which the NSC might contribute, these are most likely not accounted for.
    
\begin{figure}[t]
    	\centering
    	\includegraphics[width=0.99\columnwidth]{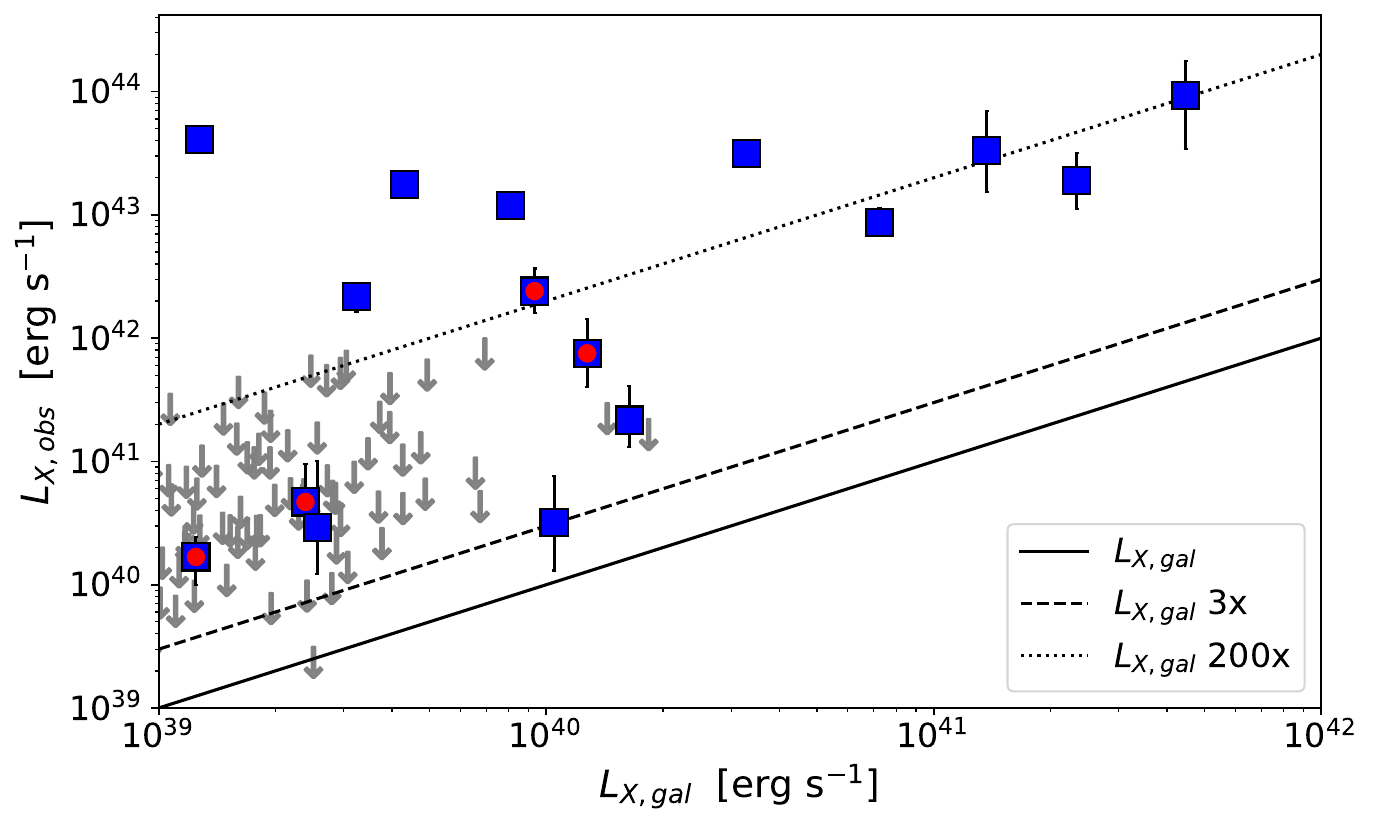}   
    	\caption{Predicted $L_{X,gal}$ \citep{Lehmer+2019:xrb} versus the observed X-ray luminosity of our eRASS:4 detected sources, both in the $2-10\,$keV range. The 1:1 relation is shown with a black solid line, while the dashed and dotted lines show the predictions increased by a factor 3 and 200, respectively. Markers containing a red circle represent new X-ray sources (see Sect.~\ref{sec:newX}). The subset of X-ray non-detected galaxies in the same range of the detected ones are shown with gray arrows, for reference.}
    \label{fig:lx_lxpred}
    \end{figure}

    In Fig.~\ref{fig:lx_lxpred}, we show the comparison between the predicted $L_{X,gal}$ and the observed X-ray luminosity of our eRASS:4 detected sources (Table~\ref{tab:results_detections}), both estimated in the $2-10\,$keV range\footnote{Therefore we do not compare these with the stacks of the soft X-ray images.}. The observed values for the detected galaxies are clearly well above the predicted ones (black solid line) including uncertainties. The dashed and dotted lines show the predictions increased by a factor 3 and 200, respectively, to guide the eye. The result of this sanity check is reassuring, since the parent sample consists of MBH candidates selected independently from UVOIR variability. This was already evident from the bottom panel of Fig.~\ref{fig:lxmstar}, although in that case the prediction for the galaxy was obtained at population level using the star formation main sequence and not individual SFR values. In the next section we discuss the role of individual luminous XRBs, relevant at the lowest end of the observed X-ray luminosity.
    
    \subsection{Contaminants: individual stellar-mass BHs}
    
    Another source of contamination comes from individual neutron stars and stellar-mass black holes at the brightest end of their luminosity function, which constitute the vast majority of the so-called ultra-luminous X-ray sources (ULXs\footnote{Here, this term is used for stellar-mass contaminants and neglects the possible presence of intermediate-mass black holes in the ULX category.}) within the host galaxies \citep[e.g., ][for a recent compilation]{Walton+2022:ULX}. Given eROSITA's point spread function ($\approx26$" half-energy width averaged over the whole field of view, \citealp{Predehl+2021:eROSITA}) we can indeed expect contamination from off-nuclear ULXs in what we have called here MBHs. However, disentangling ULXs and MBHs has revealed to be much more difficult that initially thought. As a matter of fact, recent simulations \citep{Bellovary+2021:offnuclear,Sharma+2022:hidden} and observations \citep[][but see \citealp{Sargent+2022:wandering}]{Reines+2020:wandering} have pointed out that a significant fraction of MBHs in dwarf galaxies can be displaced from the host center even up to $\sim3$\,kpc \citep{Beckmann+2023:newhorizonsimul}. Therefore, angular separation of the X-ray source from the optical nucleus alone might not be a good-enough proxy. ULXs and MBHs can be securely distinguished only if the point-like X-ray source is clearly in the outskirts of the host galaxy, or if the X-ray source is classified as a neutron star through detection of pulsations \citep[e.g.,][]{Bachetti+2014:puls} or if deep broadband spectroscopy can be carried out to distinguish between accretion states \citep[e.g.,][]{Bachetti2013:broad,Walton+2015:broad} and infer an estimate of the accretor's mass. In this work, we cross-matched our sample with the ULX catalog from \citet{Walton+2022:ULX}, which compiled \emph{XMM–Newton}, \emph{Swift-XRT}, and \emph{Chandra} data. This catalog does not overlap with the entirety of our sample, but may serve as a useful check to exclude as many known ULXs as possible. 
    Two known ULXs from \citet{Walton+2022:ULX} are within the aperture of two non-detected galaxies, whilst we found no overlap between our detected galaxies and the ULX catalog. Finally, we note that the conservative conclusion about the various stellar-mass contaminants is that at ambiguous X-ray luminosity levels $\approx10^{39}-10^{40}\,$erg\,s$^{-1}$, both the stellar-mass contaminants and MBHs are likely contributing to the total X-ray emission. This ambiguity may remain even using rich multiwavelength observations of individual nearby galaxies taken at high angular resolution \citep[e.g.,][]{Thygesen+2023:mbhvsulx}. 
    
    \subsection{New X-ray detections}
    \label{sec:newX}
    
\begin{figure*}[t]
   		\centering
   		\includegraphics[width=0.8\textwidth]{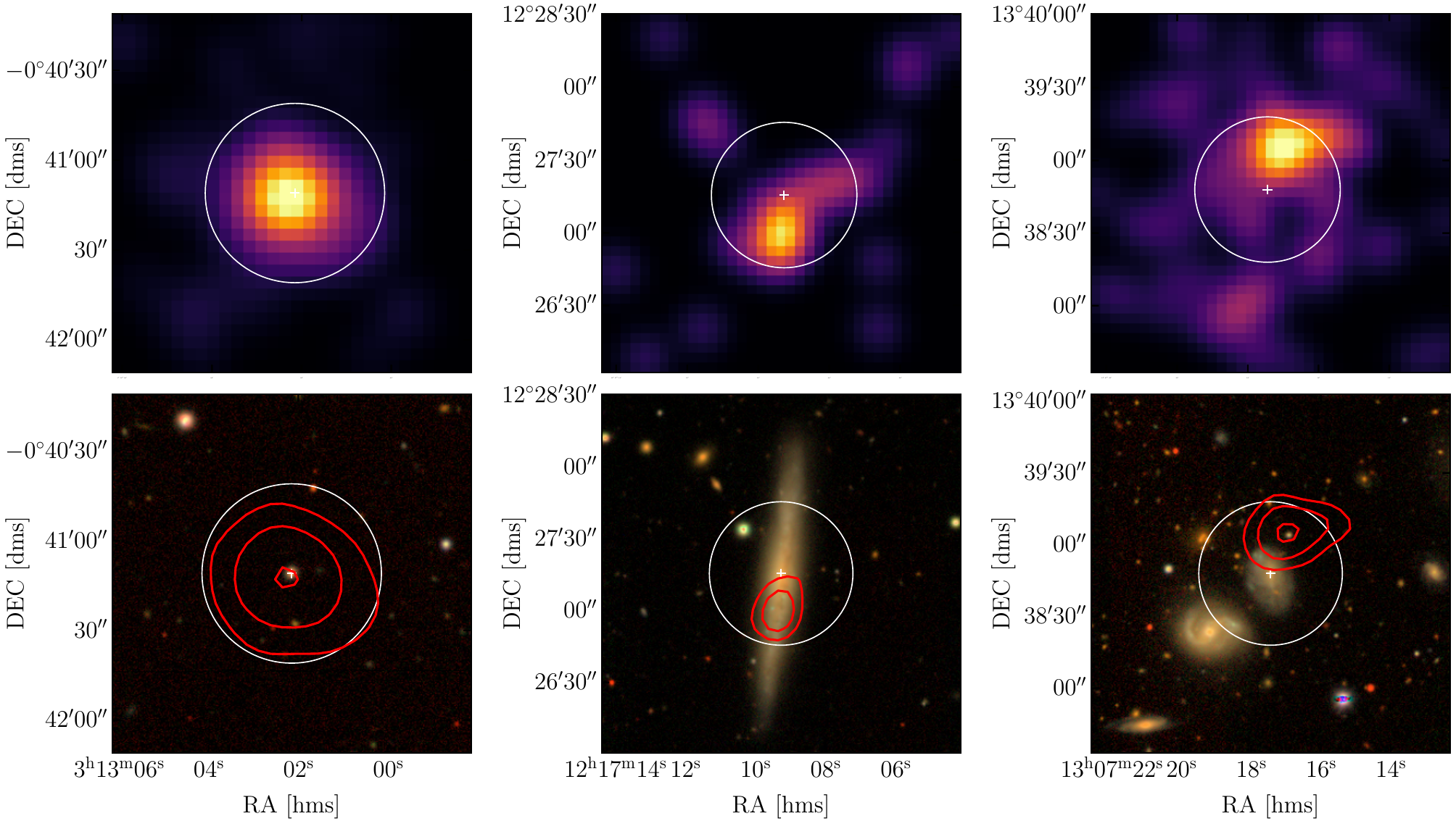}
        \caption{Optical and X-ray cutouts, as in Fig.~\ref{fig:examples}, for three of the four newly-discovered X-ray counterparts of MBH candidates, complemented by the fourth new X-ray detection shown as example in Fig.~\ref{fig:examples}. The \emph{left panels} show SDSS J031302.15-004110.9 at RA, Dec = (48.25895, -0.686379) and $z=0.131$, the \emph{middle panels} SDSS J121709.27+122714.4 (Ra, Dec: 184.28861, 12.45432) at $z=0.007$ and the \emph{right panels} SDSS J130717.44+133847.8 (Ra, Dec: 196.822679, 13.646658) at $z=0.027$, respectively. We note that the positional accuracy of the X-ray centroid is 1.5", 3.5" and 2.4" from left to right, respectively. More details on their association are presented in Sect.~\ref{sec:newX}.}
   		\label{fig:newX}
   	\end{figure*}

    We matched the 17 galaxies detected in eRASS:4 with \emph{ROSAT}, \emph{Swift-XRT}, \emph{XMM-Newton} and \emph{Chandra} catalogs in the HEASARC archives using our 30" aperture as matching radius. We have found 13 matches, all within a few arcseconds from the input coordinates. We show these matches in Table~\ref{tab:results_detections}. In the comments, we note the classification that can be inferred with a quick search on Simbad \citep{Wenger+2000:simbad}. We note the presence of two sources classified as blazars, which perhaps hints that they might be a neglected contaminant in the variable MBH searches. 
    Quite interestingly, we find that 4 of our eRASS:4 detections ($\approx25\%$) are new X-ray sources. We note that this fraction is even lower than that expected on average on the full-sky, since it is common practice to coordinate narrow-field deep multiwavelength surveys in the same sky area. This highlights the power of eROSITA with its full-sky capabilities, which balances existing and future deep pencil-beam surveys. The 4 new detections are highlighted with red circles in Fig.~\ref{fig:lx_lxpred} and their X-ray images are shown in Fig.~\ref{fig:examples} and~\ref{fig:newX}. More details are presented next.

   \subsubsection{SDSS J031302.15-004110.9 and SDSS J031743.12+001936.8}	
   	
   The first new X-ray source can be identified with SDSS J031302.15-004110.9, a known low-mass AGN at z=0.13 found to be optically-variable by \citet{Baldassare+2018:SDSS}. It is also reported as an AGN from BPT classification, with a known virial black hole mass of $\sim10^7 M_{\astrosun}$ \citep{Baldassare+2018:SDSS}. We obtained a median (and 16th, 84th percentiles) value of $L_{0.2-2.0\,keV}=43.19_{43.11}^{43.28}\,$erg\,s$^{-1}$ and a soft X-ray photon index of $\Gamma = 2.76\pm0.27$ in eRASS:4. Based on Fig.~\ref{fig:lx_lxpred}, the observed luminosity is a factor $\sim259$ above the one predicted for the cumulative XRBs in the host galaxies and it is quite extreme even for ULXs. The X-ray emission appears point-like and consistent with the optical center (Fig.~\ref{fig:newX}, left panels). We can confidently consider this source as the X-ray counterpart of the nuclear MBH. This source is present in the eRASS1 catalog \citep{Merloni+2023:erass} as 1eRASS J031302.2-004114, with (RA, Dec) = (48.25899, -0.68734) and a $1\sigma$ positional error of $2.56$".
   	
   The second new X-ray source can be associated with SDSS J031743.12+001936.8, a known low-mass AGN at z=0.069 selected from \citet{Baldassare+2018:SDSS}. This source was classified as "composite" from narrow lines diagnostics and its estimated logarithmic virial mass is $\sim6.1 \log M_{\astrosun}$ \citep{Baldassare+2018:SDSS}. We have obtained $\log L_{0.2-2.0\,keV}=42.19^{42.3}_{42.08}\,$erg\,s$^{-1}$ and X-ray photon index $\Gamma = 2.20\pm0.40$ in eRASS:4. This is the source shown in Fig.~\ref{fig:examples}, where we note a point-like X-ray emission consistent with the optical center. The observed 2-10\,keV X-ray luminosity is $\log L_{2.0-10\,keV}\sim41.88\,\log$(erg\,s$^{-1}$), a factor $\sim59$ above the luminosity predicted for the cumulative XRBs (Fig.~\ref{fig:lx_lxpred}). The optical and X-ray source coincide within 1" with the radio source FIRSTJ031743.1+001936, which has an integrated flux at 1.4 GHz of $1.82\,$mJy \citep{Helfland+2015:FIRST}. This corresponds to a luminosity density of $\log L_{1.4Ghz}\sim22.3\,$W\,Hz$^{-1}$, much brighter than the expected contribution from supernova remnants, young supernovae and ionized gas from H$_{II}$ regions \citep[e.g., see][]{Reines+2020:wandering}. Therefore we expect this to be the radio counterpart of the point-like X-ray source. We usen these estimates of X-ray and radio luminosity to infer a black hole mass through the fundamental plane of black hole accretion \citep{Merloni+2003:FP}. From the 1.4 GHz flux and assuming a flat spectrum (or, a spectrum with slope -1) in flux density units, we infer $\log L_{5Ghz}\sim39.0\,\log$(erg\,s$^{-1}$) (38.5), which yields $\log M_{BH}\sim 8.4 \log M_{\astrosun}$ (7.7). We note that the fundamental plane is only representative for radiatively inefficient black hole accretion, although it may provide us with a rough black hole mass estimate in any case. The observed luminosities are therefore too high for a stellar-mass ULX, unless its emission is beamed. While we do not know the accretion state of SDSS J031743.12+001936.8, the hard X-ray luminosity with a bolometric correction of 10 \citep{Duras+2020:Kx} corresponds to $\sim0.1 L_{Edd}$, therefore to a radiatively efficient regime. This might explain the difference between the observed mass and that predicted from the fundamental plane in the MBH scenario. Based on this, we consider this as a secure X-ray counterpart of the variable MBH. We note that this source was classified as composite based on its optical spectrum \citep{Baldassare+2018:SDSS}, which highlights once more how this selection technique is biased toward the brightest end of the MBH population. However, a closer look at the SDSS spectrum suggests the presence of a broad $H\alpha$ component that the automatic pipeline did not account for\footnote{\href{https://dr9.sdss.org/spectrumDetail?plateid=413&mjd=51929&fiber=470}{Link to SDSS spectrum}}. This source is present in the eRASS1 catalog \citep{Merloni+2023:erass} as 1eRASS J031743.0+001938, with (RA, Dec) = (49.42923, 0.32735) and a $1\sigma$ positional error of $2.82$".
      	
   \subsubsection{SDSS J121709.27+122714.4?}
   	
   	The third X-ray source is within the aperture of SDSS J121709.27+122714.4, a narrow-line galaxy at z=0.007 from \citet{Baldassare+2020:PTF}. This host is classified as star-forming using narrow line fluxes in the SDSS database\footnote{\href{https://dr9.sdss.org/spectrumDetail?plateid=1613&mjd=53115&fiber=183}{Link to SDSS spectrum}} and the narrow lines diagnostics from \citet{Kewley+2006:bpt}, adopting $\log ([OIII]/H\beta) \sim 0.25$ and $\log ([NII]/H\alpha) \sim -0.59$. From our eRASS:4 analysis, we obtained $\log L_{0.2-2.0\,keV}=39.86^{40.10}_{39.54}\,\log$(erg\,s$^{-1}$) and a hard X-ray photon index which is an unconstrained posterior with 1$\sigma$ upper limit at $\Gamma \sim 1.63$. The latter value hints for a more complex spectrum compared to a simple power-law, which will need to be explored with a deeper exposure. The detected X-ray luminosity of $\log L_{2.0-10\,keV}\sim40.22\,\log$(erg\,s$^{-1}$) is a factor $\sim13$ above that predicted for the cumulative XRBs (Fig.~\ref{fig:lx_lxpred}) and the emission is point-like (Fig.~\ref{fig:newX}, middle panels), although is consistent with being slightly off-nuclear (13'' from the optical coordinates). As discussed above, recent works have shown that MBHs in dwarf galaxies are not all coincident with the optical nucleus and the observed offset of $\sim1.9\,$kpc would be within the typical values \citep{Reines+2020:wandering,Bellovary+2021:offnuclear,Sharma+2022:hidden,Sargent+2022:wandering,Beckmann+2023:newhorizonsimul}. Nonetheless, we must consider the possibility that the X-ray-detected source is an ULX. The spectral shape would indicate that the putative ULX is in its hard ultra-luminous state \citep{Pinto+2023:ulx}, although we do not aim to state anything conclusive given the available data. Here, we note that the source is not detected in eRASS1 nor in eRASS2 and eRASS3 separately, although it is in the cumulative eRASS:3 survey at a luminosity $L_{0.2-2.0\,keV} = (4.8^{+2.0}_{-1.9})\times10^{39} \,$erg\,s$^{-1}$. It is detected in the single eRASS4 at $L_{0.2-2.0\,keV} = (1.2^{+0.6}_{-0.4})\times10^{40} \,$erg\,s$^{-1}$, hence somewhat brighter than in eRASS:3. This induces the eRASS:4 luminosity to be intermediate between the two, as reported above. No significant variability is detected within eRASS4, due to the low signal-to-noise of the individual $\sim40$\,s snapshot that eROSITA performs within the single survey \citep[e.g.,][]{Predehl+2021:eROSITA}. Overall, this might indicate that the source is variable on long (weeks to years), although not on short (hours to days), timescales. 
        
    
    \color{black}
       
    \subsubsection{SDSS J130717.44+133847.8?}

    The fourth newly-discovered X-ray source lies within the aperture around the input target SDSS J130717.44+133847.8, a galaxy at $z=0.027$ detected through infrared WISE variability \citet{Ward+2022:ZTF_Wise}. From our eRASS:4 analysis, we obtained $\log L_{0.2-2.0\,keV}=41.27^{41.37}_{41.16}\,$erg\,s$^{-1}$ and a soft X-ray photon index $\Gamma = 2.50\pm0.38$. The detected X-ray luminosity is a factor $\sim20$ above that predicted for the cumulative XRBs (Fig.~\ref{fig:lx_lxpred}). However, there is background source within the aperture at (RA, Dec) = (13:07:16.90534, +13:39:03.82002), $\sim19$" away from the input galaxy, which is coincident with the X-ray point-like source (Fig.~\ref{fig:newX}, right panels). It is identified as SDSS J130716.91+133903.8 at a Legacy Imaging Surveys photometric redshift of 1.26 \citep{Duncan+2022:desi}, and which is classified as AGN/QSO in several catalogs \citep[e.g.,][]{Richards+2015:agncat,Assef+2018:agncat} also based of its infrared ($W1-W2\sim0.8$) colors \citep{Cutri+2012:WISE}. We conclude that both the WISE variability and the eRASS:4 X-ray source are most likely attributable to the background QSO and not the foreground dwarf galaxy. In order to quantify the extent of this issue in the whole WISE-selected sample \citep{Ward+2022:ZTF_Wise}, we adopt the QSO space density to be $\sim 1.2\times 10^{-5}\,$arcsec$^{-2}$ above $W2<17.11$ for WISE AGN \citep{Assef+2013:midir,Assef+2018:agncat}. Adopting a conservative radius of three WISE pixels, each of 2.75\,arcsec in size, we would expect $\sim2.5\times 10^{-3}$ background IR-bright QSOs to be within a single WISE PSF. Therefore, we would expect $\sim200$ contaminants within the parent sample of 79879 galaxies of \citet{Ward+2022:ZTF_Wise}, which is comparable to the sample size of the 148 selected variable galaxies. However, not all the WISE QSOs are found to be variable, therefore only $\sim1.1\%$ \citep[e.g.,][]{Secrest+2020:Wise} would be detectable as contaminant in the foreground variability searches \citep[within the typical $\Delta \text{mag}\sim0.2$;][]{Ward+2022:ZTF_Wise}. Therefore the number of expected contaminants is $\sim2$ in the sample of \citet{Ward+2022:ZTF_Wise}. Since only $\sim30\%$ of their galaxies are in the eROSITA footprint, the IR source in this Section is most likely the only contaminant in the IR-selected sample. This source is present in the eRASS1 catalog \citep{Merloni+2023:erass} as 1eRASS J130716.6+133904, with (RA, Dec) = (196.81906, 13.65126) and a $1\sigma$ positional error of $4.16$".
    

    \section{X-ray undetected dwarf galaxies suggest X-ray weakness of MBHs}
    \label{sec:discussion_xweak}

    Our results find a high-fraction of non-detected dwarf galaxies with a UVOIR-variable MBHs. The typical exposure in the eRASS:4 image for the galaxies in the parent sample is only $\sim550$\,s. However, most X-ray $3\sigma$ upper limits are so deep that stacking non detected sources results in a $L_X$ estimate consistent with the predictions of the emission of the galaxy alone (bottom panel of Fig.~\ref{fig:lxmstar}). Naturally, the X-ray emission of normal galaxies and radiatively inefficient (hence low-luminosity) AGN is expected to be compatible as their relative contrast reaches unity \citep{Merloni2016:contrast}. In particular, at the same level of accretion in terms of fractions of $L_{edd}$, MBHs in dwarf galaxies are even more penalized than more massive AGN. This can be understood with order-of-magnitude scaling relations by noting that the AGN luminosities scales linearly with $M_{BH}$ for a given $L/L_{edd}$, hence $\approx M_*$ as $M_{BH}\propto M_*^{\beta}$ with $\beta \approx 1$ or larger \citep{Reines+2015:coev}, whilst the galaxy luminosity scales linearly with SFR, which in turns scales as $SFR\propto M_*^{0.7}$ at $z=0$ for main sequence galaxies \citep{Whitaker+2012:sfrms}, ignoring redshift and metallicity dependencies for simplicity. As a matter of fact, we have already showed this with order-of-magnitude predictions in the bottom panel of Fig.~\ref{fig:lxmstar} with black shaded contours and the dotted red line, which are related to normal galaxies and inefficient AGN accreting at $\sim 10^{-3} L_{edd}$, respectively. Therefore, at this stage we can only conclude that the X-ray luminosity from the stacked non-detected sources is compatible with both.

    \begin{figure*}[t]
    \centering
    \includegraphics[width=0.75\textwidth]{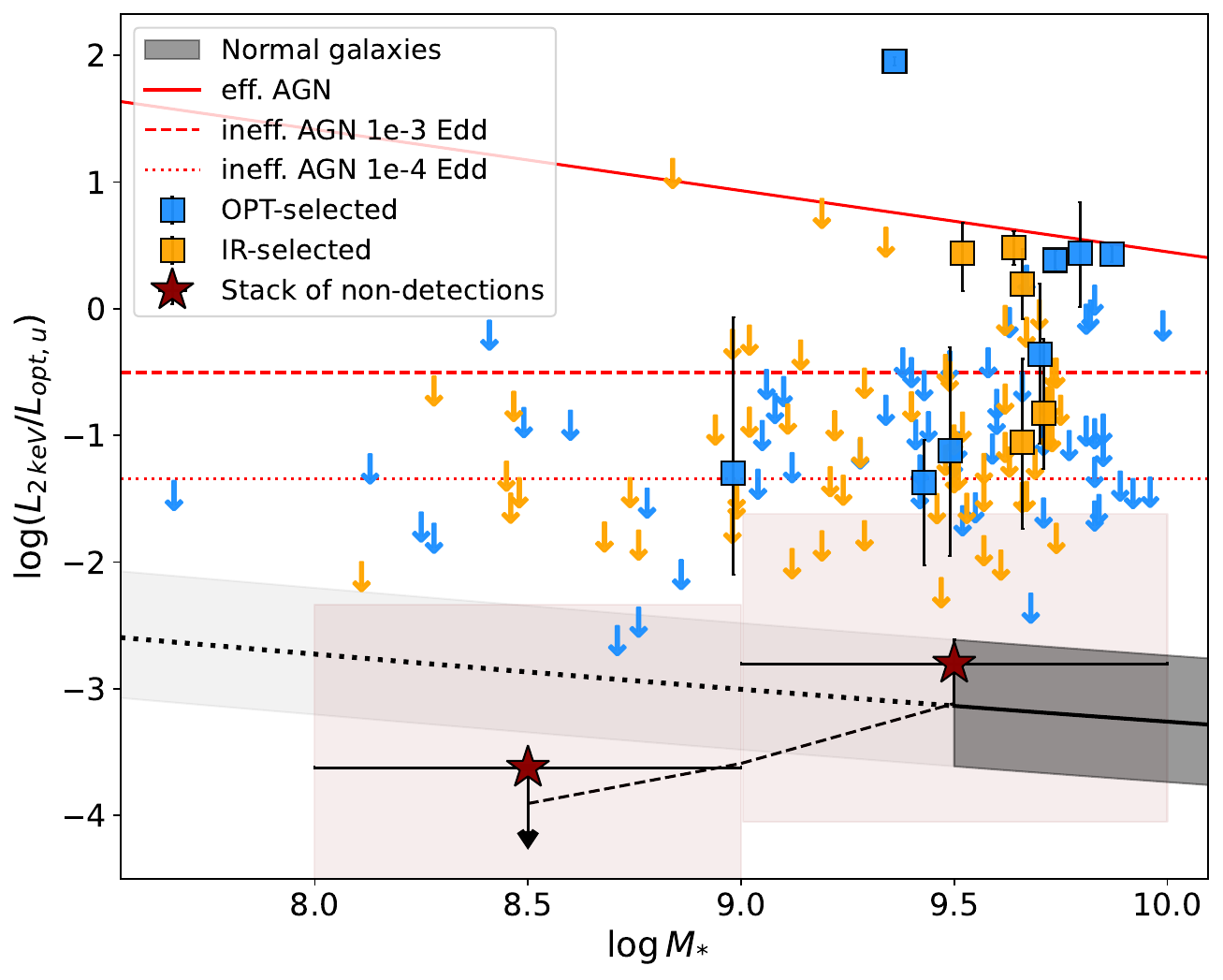}  
    \caption{Observed X-ray to optical ratio as a function of galaxy stellar mass. Squares indicate X-ray detections and arrows 3$\sigma$ upper limits, color-coded by the waveband used for variability selection. The luminosity from the stacked images of non-detected galaxies are shown with red stars (and their uncertainties with shaded contours, see the text in Sec.~\ref{sec:discussion_xweak}). The black contour indicate the predicted X/O from normal galaxies \citep{Lehmer+2019:xrb}, the dotted black line its extrapolation and the dashed black line a tentative correction for the low-mass end \citep[][and refer to Sect.~\ref{sec:mstar} in this work]{Gilfanov+2004:xrbs,Lehmer+2019:xrb}. Red lines show the predicted X/O for AGN in their radiatively efficient phase \citep[solid red line;][]{Arcodia2019:lxluv}, compared to inefficient ones at $\sim10^{-3}$ or $\sim10^{-4} L_{edd}$ \citep[dashed and dotted red lines;][]{Merloni2016:contrast,Ruan+2019:aox}.
    }
    \label{fig:lxlo_mstar}
    \end{figure*}

    However, we can gain more information from the SED adding the information from the optical band in the picture. In particular, we know the brightness of these galactic nuclei (Fig.~\ref{fig:hist}) and we can attempt to use typical X-ray-to-optical (X/O) scaling relations to put our observations into a wider context. We highlight this in Fig.~\ref{fig:lxlo_mstar}, where we show observed X/O luminosity ratios as a function of stellar mass for all the MBHs below $z=0.1$. Squares represent detections within eRASS:4, arrows are $3\sigma$ upper limits. Both are color-coded based on the variability selection between optical (blue) and infrared (orange), to highlight the lack of obvious biases in either. The u-band flux (\texttt{EL\_PETRO\_FLUX}) is obtained from the parent SDSS NASA-Sloan Atlas sample. We add X/O values computed from the stacked non-detections as follows. The monochromatic rest-frame 2\,keV luminosity is obtained dividing the stacked luminosity between $0.5-2.0\,$keV (see Sect.~\ref{sec:xray} and the bottom panel of Fig.~\ref{fig:lxmstar}) by a conversion factor obtained from the detected galaxies (e.g., the squares in Fig.~\ref{fig:lxlo_mstar}), taking the median value of their observed $F_{2\,keV}/F_{0.5-2.0\,keV}$ ratio. The optical luminosity (their uncertainty) for the stacked value is obtained using the median (1st and 99th percentiles) of the observed u-band flux within the two stellar-mass bins. The statistical uncertainties from the stacks are shown with vertical errorbars as in Fig.~\ref{fig:lxmstar}, whilst the uncertainty coming from the range of u-band used for computing the stacks' X/O is shown with a darkred contour. 
    
    Observed $\log (L_{2keV}/L_{opt,u})$ are compared with predictions from models of normal galaxies (gray contour) and AGN (red lines). For normal galaxies we used scaling relations from \citet{Lehmer+2019:xrb} and \citet{Merloni2016:contrast}, using the star formation main sequence \citep{Whitaker+2012:sfrms} and mass-to-light ratios between 1-10. As explained in Sect.~\ref{sec:mstar}, the galaxy predictions are calibrated only at the high-mass end, and we show with a dotted black line the extrapolation, whilst we attempt to correct for underpopulated low-mass and low-SFR galaxies \citep{Gilfanov+2004:xrbs,Lehmer+2019:xrb} drawing the dashed black line. For AGN, we computed the optical luminosity following \citet{Merloni2016:contrast} and the X-ray luminosity from the $L_X-L_{UV}$ relation for radiatively-efficient \citep{Arcodia2019:lxluv} and -inefficient \citep{Ruan+2019:aox} AGN. The former prediction is shown with a red solid line, the latter with dashed (dotted) for inefficient accretion at $\sim10^{-3}$ ($\sim10^{-4}$) of $L_{edd}$. We confirm that, as in Fig.~\ref{fig:lxmstar}, the stacks are compatible with the emission of normal galaxies. Since the u-band filter has an effective wavelength at $\sim3565\AA$, whilst these scaling relations are calibrated at $\sim2500\AA$ or $\sim3000\AA$ \citep{Arcodia2019:lxluv,Ruan+2019:aox}, we also computed the X/O ratios using GALEX's near-UV filter at $\sim2300\AA$ (Fig.~\ref{fig:lxlo_galex}). The comparison between observed X/O and model predictions remains qualitatively the same and in fact using GALEX even fainter X/O values are obtained. Therefore, the observed X-ray weakness is even more enhanced compared to the bottom panel of Fig.~\ref{fig:lxmstar}, once the optical/UV luminosities are used to provide a characteristic SED shape. The underlying assumption is that the host galaxy is contaminating, but not dominating the optical emission, which is reasonable given that the MBH has to contribute enough to the flux to allow the inference of its presence through variability, at least in the cases of moderate $\Delta_{mag}$. Furthermore, the typical SED of the MBH candidates does not seem to show worryingly or ubiquitously dominant contributions from the stellar component alone \citep{Burke+2022:DES}, specially for the bluer optical and UV filters used here. We also indirectly quantified the impact of the host galaxy contamination in the optical band by separating star-forming galaxies from AGN, classified based on narrow lines diagnostics \citep{Baldwin+1981:bpt}, using several different classification methods (see Appendix~\ref{sec:moreplots} and Fig.~\ref{fig:bpt}). We obtain that there is no significant difference in X-ray luminosity and stellar-mass between these two categories, implying that we are not biased toward X-ray detections only for galaxies with a strong central ionizing source inferred from the optical photometry or spectroscopy. Finally, the X/O predictions from AGN at low Eddington ratios are also, to some extent, contaminated by the galaxy in the optical-UV band \citep[e.g.,][]{Ruan+2019:aox}, validating our comparison in Fig.~\ref{fig:lxlo_mstar}. We conclude that canonical AGN disk-corona SEDs \citep[e.g.,][]{Arcodia2019:lxluv,Ruan+2019:aox} would predict the X-ray emission from the MBHs in these galaxies to be much brighter than observed, even for predictions of low-luminosity AGN \citep{Ruan+2019:aox}.

    
    We note that the possible X-ray weakness of MBHs in dwarf galaxies, or their unusual SEDs, compared to more massive AGN was reported before for a few of cases \citep{Dong+2012:xweak,Simmonds+2016:xweak,Baldassare+2017:xuv,Cann+2020:xweak,Burke+2021:weak,Gueltekin+2022:fpimbh,Urquhart+2022:lowedd,Messick+2023:ovar}, although this is the first confirmation on a large sample of fairly homogeneous X-ray exposures of dwarf galaxies. The optical variability selection in these galaxies (directly or through the infrared echo) is thought to indicate the presence of a variable radiatively-efficient AGN accretion disk \citep{Burke+2021:tdamp}, whilst the X-ray upper limits and stacked X-ray images obtained in this work are, at best, compatible with AGN accreting at $\sim10^{-3}-10^{-4} L_{edd}$ and, at worst, consistent with and inactive or absent black hole. This begs the question of whether these two observables, UVOIR stochastic variability and X-ray data, are consistent. Before analyzing the possible physical interpretation and consequences, we briefly discuss possible biases that might cause MBHs to appear unusually X-ray weak (Sect.~\ref{sec:biases}). We stress again that, in order to avoid strong redshift effects and to be consistent with the sources used for the X-ray stacking analysis, we limit the discussion to the 134 sources with X-ray products in eRASS:4 which are below z<0.1.

    \section{On the possible biases for the observed X-ray weakness}
    \label{sec:biases}
    
    First, we do not find any obvious correlation between X-ray (non-) detection and variability significance from the parent samples. For instance, among the galaxies \citet{Burke+2022:DES} we have only detected the one with highest and the one with lowest variability significance, and the four detected galaxies from \citet{Baldassare+2020:PTF} are also homogeneously distributed in terms of variability significance. Furthermore, we investigated in Appendix~\ref{sec:moreplots} whether the observed X-ray weakness depends on the variability significance, both for optically- \citep[e.g.,][]{Baldassare+2018:SDSS,Baldassare+2020:PTF} and IR-selected \citep[e.g.,][]{Ward+2022:ZTF_Wise} variable galaxies. For the optically selected variable galaxies, we also investigate the dependence on the number of data points in the optical light curve or the total baseline. We show this in Fig.~\ref{fig:lxlo_sigmavar} and~\ref{fig:lxlo_irsign} and no significance trend is evident. For the optically selected variable galaxies, we also stacked lower- and higher-significance sources from \citep{Baldassare+2020:PTF} in the $\log M_*=9-10$ bin separately and obtained no significant difference, although we found weak evidence indicating that the stacked image on the higher-significance galaxies contained brighter signal (see Appendix~\ref{sec:moreplots}). Furthermore, we tested whether the observed X-ray weakness depends on the optical classification from narrow-lines diagnostics \citep{Baldwin+1981:bpt} using several techniques, and we found again no obvious difference (Fig.~\ref{fig:bpt} and Appendix~\ref{sec:moreplots}). However, formally our X-ray observations did not confirm the nature of most of these MBHs as such. From X-rays alone, a possibility is that these galaxies would be mostly inactive and lack significant accretion all-together. Hence, a conservative possibility that we must consider is that the bulk of the variability-selected MBHs is contaminated, as also a bias spread to most of the light curves, regardless on the inferred variability significance, would appear uncorrelated with the X-ray non-detections. This is very unlikely, although it is still relevant to discuss possible known contaminants. Possible spurious sources within the methodology typically adopted to select variable AGN \citep[e.g.,][]{Butler+2011:sigma,Burke+2022:DES} could be long-lived stellar transients or variables \citep[e.g.,][]{Burke+2020:long-lived,Burke+2021:weak,Kokubo+2022:lbv,Rizzo-Smith+2023:sne}, although they are expected to contaminate the selected MBHs in small numbers. Another contaminating component which is nearly ubiquitous in these galaxies in the NSC, although its old stellar population is not expected to imprint any variability \citep{Neumayer2020:NSC}. Therefore, for any bias in the optical photometry to impact our systematic X-ray weakness, it would have to be currently unknown and worryingly extended to the bulk of the parent galaxy samples. It is worth mentioning that, despite the large overlap in the parent sample of dwarf galaxies, variability studies using data from the Palomar Transient Factory \citep[e.g.,][]{Baldassare+2020:PTF} and the Zwicky Transient Facility \citep[e.g.,][]{Ward+2022:ZTF_Wise} have limited overlap in their respective MBHs candidates. In particular, $\sim11\%$ of the ZTF candidates were selected also by PTF, and, viceversa, only $\sim3\%$ of the PTF candidates were also selected by ZTF \citep{Ward+2022:ZTF_Wise}. However, the possible origin of this discrepancy may lie in the difference cadence, scatter and total baseline of data obtained with PTF and ZTF. In particular, PTF has median baseline in the parent sample of $\sim4\,$yr, reaching higher detection fractions for galaxies with baseline up to $\sim6-7\,$yr \citep{Baldassare+2020:PTF}, while ZTF data have a typical baseline of $\sim3\,$yr. Therefore, it is possible that the MBHs selected by PTF and missed by ZTF were mostly variable on timescales comparable with or longer than the ZTF baseline. This would be supported by the fact that the 5 in common have variability power at much higher timescales compared to the rest of ZTF-selected MBHs. Conversely, the ZTF-detected MBHs might have been missed by PTF due to its reduced sensitivity to variability over the timescale of months, compared to ZTF. Therefore, as much as some of the variable MBHs might be spurious sources (i.e. normal galaxies with a dormant black hole or no black hole all-together), this is unlikely to be the case for most of the 121 undetected X-ray MBHs of the low-z sample \citep[as also discussed in][albeit with a much smaller sample]{Messick+2023:ovar}. Without dedicated simulations quantifying the purity and completeness of the variability searches, we are unable to identify a subset of secure MBHs or to quantify the spurious fraction in our sample.
    
            
    Furthermore, \citet{Baldassare+2017:xuv} noted a lower X/O in their eight broad-line MBHs and discuss that enhanced nuclear star formation might be a contaminant to their optical-UV data. In our sample, the optical nucleus would have to be dominated by the galaxy to the extent of altering X/O, but not to the extent of impeding the detection of AGN-like optical variability on top of the galaxy continuum, which requires suspicious fine tuning of the ratio between AGN and galaxy in the optical, considering the several tens of X-ray weak sources found here. In \citet{Simmonds+2016:xweak}, it was noted that X/O variability and non-simultaneity would scatter the X-ray estimates toward both the brighter and fainter direction and not systematically toward the latter. We confirm this by cross-matching the eROSITA estimates with the fourth \emph{XMM-Newton} serendipitous source catalog \citep{Webb+2020:xmmdr12} and the \emph{Chandra} Source Catalog \citep{Evans+2020:chandra}. We show in Fig.~\ref{fig:fxfx} the resulting comparison, which shows compatible fluxes between the eROSITA, \emph{XMM-Newton} and \emph{Chandra} across the different epochs. As a consequence, since there is no evidence of any long-term variability effect between the X-ray epochs, it is unlikely that the X/O weakness is solely due to long-term variability.
    
    The possible role of X-ray absorption needs to be assessed, as it surely impacts some of these galactic nuclei. Using the observed WISE magnitudes and X-ray upper limits, we can put a $3\sigma$ lower-limit prediction on the $N_H$ (cm$^{-2}$) required for these nuclei to be obscured, under the assumptions that they follow multiwavelength prescriptions of more massive obscured AGN. Using the relation between $N_H$, X-ray luminosity and W3 magnitude from \citet{Asmus+2015:mir}, the median lower-limit is $\log (N_H/$cm$^{-2})>23.6$. This implies that the typical dwarf galaxy in our sample would need to be Compton thick. In general, it is true that in the most extreme case $\approx50\%$ of the existing nuclear BHs are Compton thick \citep[e.g.,][for a recent work]{Carroll+2023:cthick}. However, the MBHs in this study are not simply randomly-selected low-mass galaxies for which this statistics may apply. They were selected through UVOIR variability, which therefore excludes that the SED is heavily obscured. Therefore, the observed X-ray weakness is unlikely to be due to extreme obscuration. Since our sample contains also IR-selected objects, let us still pessimistically assume that all IR-variable MBHs are X-ray obscured. One would still need to account for the remaining optically unobscured nuclei. Moreover, we observed X-ray weakness homogeneously between optically- and infrared-variable MBHs, which argues against systematic obscuration in all the nuclei of these dwarf galaxies. As a matter of fact, we stacked the X-ray images of the non-detected IR-selected and optically selected galaxies separately in the $\log M_*=9-10$ bin and found compatible results and even weak evidence that the X-ray signal of the stacked IR-selected galaxies is brighter than the optically selected, which would argue against wide-spread obscuration in the latter. In particular, using as background estimate the median signal between $15-50\,$kpc (see Sect.~\ref{sec:xray}), we obtain a median value of $L_{0.5-2.0\,keV}=(1.0\pm0.9)\times 10^{39}\,$erg\,s$^{-1}$ and $(1.0\pm0.7)\times 10^{39}\,$erg\,s$^{-1}$, for optically- and IR-selected non-detected MBHs, respectively. Instead, conservatively using as background estimate the 84th percentile of the signal between $15-50\,$kpc the optically selected galaxies are non-detected at $L_{0.5-2.0\,keV}<1.6\times 10^{39}\,$erg\,s$^{-1}$, whilst the IR-selected ones are still detected at $(7.3\pm6.9)\times 10^{38}\,$erg\,s$^{-1}$. Hence, X-ray obscuration is not considered to play a major role in the observed X-ray weakness.
    
    We conclude that it is likely that only some of the galaxies in our sample might suffer from one or more of the above-mentioned effects (spurious trigger in the variability searches, X-ray variability and X-ray absorption). The only way for biases to be extended to the whole sample studied here, would imply that most IR-selected MBHs are Compton thick and that most of the optically selected are systematically flawed by currently-unknown physical, instrumental or statistical contaminants. Arguably, this seems quite unlikely. Therefore, we discuss possible physical interpretations for the observed X-ray weakness in MBHs in dwarf galaxies.

    \section{On the possible physical interpretations for the observed X-ray weakness}

    We generically refer to a canonical corona \citep[e.g.,][]{Haardt+1991:corona} as a magnetically-powered plasma in the immediate vicinity of the black hole, with electrons kept hot and accelerated with a high duty cycle \citep[e.g.,][]{Balbus+1991:corona,DiMatteo+1998:corona,Beloborodov+2017:corona,Zhang+2023:reconn}. Its emission typically scales with the optical-UV emission for radiatively-efficient BHs \citep{Arcodia2019:lxluv} and with radio for the inefficient ones \citep{Merloni+2003:FP}. To summarize the intents of this section, in this work we have obtained that the majority or UVOIR-variable MBHs are X-ray weak, with luminosity similar to those of normal galaxies. In Sect.~\ref{sec:biases} we controlled for potential biases, and excluded X-ray obscuration as a systematic contaminant. Under the assumption that UVOIR variability is a robust method that traces some level of accretion in these nuclei (be it radiatively-efficient or -inefficient), the central MBH must be active to some degree. Even for low Eddington ratios X-rays are expected and are, in fact, a significant or dominant contribution in the bolometric SED compared to optical and UV proxies \citep{Merloni+2003:FP,Kubota+2018:sed,Arcodia2019:lxluv}. Hence, here we discuss possible physical interpretations, which would be due to a different behavior present in low-mass nuclei compared to more massive ones: for instance, in a different structure or powering of the accretion disk-corona system, different fueling of gas and magnetic field toward the galaxy nucleus, or a different variability behavior.
    
    We start discussing the case in which the UVOIR variability is uniquely tracing temperature fluctuations in a radiatively-efficient accretion disk \citep{Burke+2021:tdamp}, then the observed X-ray weakness compared to the optical would suggest that active MBHs do not follow standard AGN accretion SEDs or X/O values (e.g. see Fig.~\ref{fig:lxlo_mstar}). 
    Interestingly, in newborn (hence not accumulated secularly) accretion flows following tidal disruption events and quasi-periodic eruptions, which are observed in the same low-mass regime of the black hole and galaxy populations too \citep{Wevers+2017:tdes,Wevers+2022:qpes}, the hard X-ray corona is usually missing \citep[e.g.,][]{Miniutti+2019:qpe1,Saxton+2020:XTDES,Giustini+2020:qpe2,Arcodia+2021:eroqpes,Mummery+2023:tdes}. However, if the lack of a canonical corona were to be the only cause of the X-ray weakness, then one would still expect to detect more of these MBHs by detecting the tail of the radiatively-efficient disk emission in the soft X-rays (where eROSITA is most sensitive), which is expected to be observable from these putative $\sim10^{5}-10^{6.5} M_{\odot}$ black holes and it is, in fact, seen for the above-mentioned transients.

    Another option is that optical/IR variability searches would trigger not only stochastic variability from the thermal emission of a radiatively-efficient accretion disk \citep{Burke+2021:tdamp}, but also variability from the nonthermal SED of radiatively-inefficient ones. This is most evident in the submillimiter \citep{Chen+2023:submm}, but its SED extends to higher frequencies too \citep[e.g.,][]{Yu2011:uvlum,Mason2013:IRadaf,Nemmen+2014:adaf,Fernandez+2023:LLAGN}. In this case, no tail of the accretion disk emission is expected in the soft X-rays, therefore one needs to worry solely about the possible absence of a corona. For these radiatively-inefficient MBHs, one would expect the X-rays to align with X/O predictions of such accretion regimes and, most importantly, with radio estimates along the fundamental plane of black hole accretion \citep{Merloni+2003:FP}. However, neither the former (dashed and dotted red lines in Fig.~\ref{fig:lxlo_mstar}) nor the latter (Fig.~\ref{fig:fp}) is observed. In particular, in Appendix~\ref{sec:moreplots} and Fig.~\ref{fig:fp} we show that, despite the low sample statistics of sources with an archival radio flux above the SFR estimate, these MBHs are X-ray weak even in the fundamental plane. This is at odds with the interpretation that the observed X-ray weakness is merely due to the low-luminosity nature of these MBHs. We note that we used standard scaling relations with stellar mass \citep{Reines+2015:coev} to obtain the black hole mass. In principle, if these black holes were overmassive with respect to their stellar masses, this would not only alleviate the tension with the fundamental plane, but also explain why we do not see the exponential tail of the accretion disk emission in the soft X-rays. However, since even the $3\sigma$ upper limit values are off by at least $\sim1-1.5\,$dex from the mean fundamental plane (Fig.~\ref{fig:fp}), one would need to offset the black hole mass by at least $\sim1.3-1.9\,$dex \citep[given the 0.78 dependence from $\log M_{BH}$;][]{Merloni+2003:FP}, which is quite extreme. Further, we note that the observed X-ray weakness in the fundamental plane is consistent with other results in the literature \citep{Gueltekin+2022:fpimbh}, albeit still with low sample statistics. If confirmed in the future with wide area survey matches between X-rays, such as eROSITA \citep{Predehl+2021:eROSITA}, and radio, such as ASKAP-EMU \citep{Norris+2011:EMU}, this would indeed imply that, at least in UVOIR-variable MBHs, X-rays are decoupled from both optical and radio, compared to standard accretion modes at other black hole masses. 
    
    An intriguing option is that a significant fraction of MBHs in dwarf galaxies is spoon-fed by transient accretion events, e.g. by tidal disruption events \citep[e.g., see][]{Zubovas+2019:mbhtde,Baldassare+2022:nsc,Messick+2023:ovar}. In this case a corona is not necessarily expected and even if standard SEDs are seen in TDEs too \citep[e.g.,][]{Wevers+2020:tdes}, their complex multiwavelength signatures surely do not follow standard AGN scaling relations at all times. For instance, a case-study of the possible intermittent activity in these galactic nuclei is the possible short-lived ($<1.6$\,yr) flare that is thought to have recently happened ($\approx200\,$yr ago) in the nucleus of the Milky Way \citep{Marin+2023:sgrflare}. 
    However, the UVOIR variability was observed to be stochastic, non-transient and selected with baselines longer than the typical nuclear transient duration, and transient emission is normally excluded from these studies \citep[e.g.,][]{Baldassare+2018:SDSS, Baldassare+2020:PTF, Burke+2022:DES, Ward+2022:ZTF_Wise}. As much as unusually long-lived transients may contaminate some individual galaxies, it is unlikely that this contaminant is present in tens-hundreds of galaxies. More fundamentally, it would imply that TDEs are much more common than what both observations and theory suggest \citep[e.g.,][]{Vanvelzen+2020:rates}. 
    Alternatively, it is possible that MBHs in low-mass galaxies are typically powered with a much lower duty-cycle compared to more massive nuclei. Intriguingly, a low-luminosity analog with a lower duty cycle in X-rays compared to more frequent activity in the optical and infrared is Sgr\,A*. This is not an unreasonable example since the SED of Sgr\,A* is, for instance, compatible with that of M81, which is about four orders of magnitudes brighter \citep{Markoff+2008:M81}. The infrared variability of Sgr\,A* (and we assume, by extension, its optical too) appears stochastic with a red noise character \citep{Witzel+2018:pow,Gravity2020:flux}. Conversely, Sgr\,A* shows flares in the X-ray band for only $\sim2\%$ of the time, considering roughly a flare a day lasting $\sim30\,$min \citep{Neilsen+2013:sgr,Ponti+2015:sgr,vonFellenberg2023:flareduration}. If this behavior were to happen in galaxies such as those in our parent sample, albeit at much higher luminosity compared to Sgr\,A*, it would potentially trigger stochastic random walk variability searches within the typical light curve cadences \citep[e.g., see][]{Baldassare+2020:PTF,Ward+2022:ZTF_Wise}, considering the red noise character of the IR light curve. On the other hand, in the X-ray band there would be a very high likelihood of catching the source in the quiescent state, therefore the OIR-variable galaxy would appear undetected in X-rays. However, a low-duty cycle is generally unlikely to explain the ubiquitous X-ray weakness we observe, since eROSITA and archival XMM-Newton/Chandra X-ray fluxes, taken at different epochs separated by years, align quite nicely for the few sources in common (Fig.~\ref{fig:fxfx}). Therefore, it would be quite unlikely to have the putative low duty-cycle impacting the X/O and X/radio ratios only, and not the X-ray versus X-rays long-term comparisons.

    Hence, we discuss a possible physical picture for our UVOIR-variability selected MBHs. UVOIR variability is likely tracing both thermal and nonthermal processes \citep[e.g.,][for the latter case]{Igumenshchev+1999:adafvar,Fernandez+2023:LLAGN} in the accretion flow, depending on the accretion rate of the source. Thus, the MBHs found through these variability searches can be both radiatively-efficient and -inefficient \citep[e.g.,][]{Yu2011:uvlum,Fernandez+2023:LLAGN}, depending on the overall luminosity and SED (Fig.~\ref{fig:lxlo_mstar}). The fainter accretion regime is unsurprisingly more common \citep[e.g.,][for more massive galaxies and AGN]{Aird+2012:prob,Bongiorno+2012:prob,Georgakakis+2017:prob}, hence the high number of non-detected MBHs in dwarf galaxies, which are also predicted to be dominant from simulations \citep{Sharma+2022:hidden}. For these inefficient MBHs, radio traces their synchrotron continuum as expected, forming a nuclear SED to which X-rays should contribute too \citep[e.g.,][]{Fernandez+2023:LLAGN}, were these MBHs to follow standard scaling relations valid at other black hole masses \citep{Merloni+2003:FP}, but somehow they do not seem to be (e.g., Fig.~\ref{fig:fp}). Hence, the X-rays are weak compared to both efficient (i.e. optically-bright) and inefficient (i.e. radio-bright) accreting MBHs. Therefore, it would seem natural to conclude that a canonical X-ray corona might be missing in the bulk of the MBH population in dwarf galaxies all-together. As much as there is general agreement that the X-ray corona is magnetically powered, the formation mechanism of this highly magnetized coronal region is still unsolved \citep[e.g.,][]{Sironi+2020:corona,ElMellah+2022:corona}. This likely requires that gas with a large magnetic field is funneled toward the black hole \citep[e.g.,][and references therein]{Begelman+2023:magn}. This is a highly uncertain and understudied field, but we may interpret our observational result as follows, namely that MBHs in dwarf galaxies are not as efficient as more massive ones in sustaining a magnetically-powered corona. Under the assumption that the magnetization of the corona and that of the large-scale gas feeding the black hole are somehow linked, this means that the strength and order of the magnetic field in the nuclei of low-mass galaxies is less effective, compared to more massive galaxies and nuclei \citep[e.g., see][]{Begelman+2023:magn}.
    
    We now outline a few major differences between low-mass and massive galaxies. As a matter of fact, dwarf galaxies have a much shallower nuclear potential well which might cause the lack of a clear galactic center all-together \citep{Bellovary+2021:offnuclear} and observations of compact dwarf galaxies indeed clearly show a rather clumpy and inhomogeneous interstellar medium \citep[e.g.,][]{Cairos+2001:bcd,Cairos+2009:starb,James+2020:inhom,Cairos+2021:musedwarf,Kimbro+2021:dwme}. Furthermore, dwarf galaxy mergers do not seem to funnel gas toward the nucleus as efficiently as in more massive mergers \citep[e.g.,][]{Privon+2017:funnel} and morphological studies indicate major mergers are rarer at the low-mass end \citep[e.g.,][]{Casteels+2014:gamamerg,Guzman+2023:merg}. Another major difference between low- and high-mass galaxies is the high fraction of nuclear star clusters in the former and the lack thereof in the latter. Indeed, NSCs are thought to be directly linked to the growth of the MBH \citep[e.g.,][]{Kritos+2022:nscBH}. Whether (and how) all the above-mentioned differences eventually impact the formation and powering of the X-ray corona (still, in general, an open question) at $\sim10$ gravitational radii remains to be established. We invoke further study on the magnetization of galaxies of different masses and their connection with the channeling of gas toward the central regions of the galaxy down to the black holes. Until then, the scenario discussed here is merely a tantalizing possibility which can not be quantitatively supported. 

    

    \section{Summary and future prospects}

    The search for MBHs ($M_{BH}\approx10^4-10^6 M_{\astrosun}$) in the nuclei of low-mass galaxies ($M_{*}\lessapprox10^{10} M_{\astrosun}$) is of paramount importance to constrain black holes seeding and their growth over time, although it is a challenging task (e.g. see \citealp{Greene+2020:imbhs} for a recent review). A promising way to find MBHs at lower luminosity, compared to searches based on broad and narrow optical lines, was provided by the growing number of high-cadence photometric surveys which allow selection of MBHs through UVOIR variability. 
    In this less efficient accretion regime, X-ray and radio searches are also particularly useful in finding and confirming low-luminosity MBHs, although these observations have been so far limited to small samples. 
    This is where eROSITA \citep{Predehl+2021:eROSITA} comes into play with its homogeneous all-sky survey and its selection function calibrated with simulations \citep[e.g.,][]{Seppi+2022:erass1_simul}. It is also common practice, when there is not an a priori knowledge on the presence of a MBH in the nucleus, to study subsamples of galaxies with multiwavelength detections across the SED. However, this approach is naturally limited in studying a biased selection of active MBHs with canonical SEDs. Ultimately, it is still unclear to what extent selection techniques from different wavebands compare with one-another at the fainter end of accretion. 

    In this work, we presented the first large systematic investigation of the X-ray properties of a sample of known MBH candidates, which has the advantage of providing a sample with occupation and active fraction of one. We focused on MBHs selected through UVOIR variability (Sect.~\ref{sec:sample} and Fig.~\ref{fig:hist}). In Sect.~\ref{sec:xray}, we extracted X-ray photometry and spectra (e.g., Fig.~\ref{fig:examples}) of a sample of 214 (208) UVOIR variability-selected MBHs from the eRASS1 (eRASS:4) image and significantly detect 11 (17) of them, hence $5.1_{-1.5}^{+2.1}\%$ ($8.2_{-2.0}^{+2.5}\%$; Sect.~\ref{sec:results}). The detection fraction mildly increases with the stellar mass of the galaxy (bottom left panel of Fig.~\ref{fig:detfrac}) and so does the observed X-ray luminosity (Fig.~\ref{fig:lxmstar}). We present a summary of our sample and the X-ray results in Table~\ref{tab:results}. Out of the 17 detected galaxies from the deeper eRASS:4 image, 4 are newly-discovered X-ray sources (Table~\ref{tab:results_detections} and Fig.~\ref{fig:examples} and~\ref{fig:newX}), two of which are securely X-ray counterparts of the variable MBHs, whilst the other two remain ambiguous (Sect.~\ref{sec:newX}).

    For the first time on a large ($\sim200$) number of galaxies, we dedicate  significant attention to the many of them which are undetected in X-rays (Sect.~\ref{sec:discussion_xweak}). The eROSITA survey is shallow (e.g. the median net exposure for this sample is $\sim550$\,s in eRASS:4), although its selection function as a function of X-ray flux is well-calibrated from all-sky simulations \citep[][and top left panel of Fig.~\ref{fig:detfrac}]{Seppi+2022:erass1_simul}. Most importantly, stacking the images of non detected sources results in a $L_X$ estimate which is orders of magnitudes fainter than the X-ray detections, and consistent with the predictions of the emission of the galaxy alone (bottom panel of Fig.~\ref{fig:lxmstar}). In particular, no X-ray signal is detected in the stacked images below $\log M_*=9$. However, the X-ray emission of normal galaxies and radiatively-inefficient, hence low-luminosity, AGN becomes notoriously indistinguishable, specially if it is unresolved. Nonetheless, the advantage of the parent sample being composed by known MBHs from UVOIR-variability is to exclude that these MBHs are overall intrinsically faint. Therefore, their X-ray weakness in comparison with their UVOIR variability is puzzling. In particular, we investigate that most X-ray $3\sigma$ upper limits are so deep that they lie well below the predictions based on more massive AGN, both for radiatively-efficient (comparing X-rays with predictions from optical proxies, Fig.~\ref{fig:lxlo_mstar}) and -inefficient ones (comparing with radio proxies, Fig.~\ref{fig:fp}). However, X/O comparisons are surely contaminated by the galaxy and future work will need to reproduce this analysis decomposing the AGN contribution from the optical-UV magnitudes used (Fig.~\ref{fig:lxlo_mstar} and~\ref{fig:lxlo_galex}), and X/radio comparisons in this work are limited by much lower statistics (Fig.~\ref{fig:fp}) and will need to be assessed with larger radio samples.

    We carefully considered potential biases which would cause the observed X-ray weakness to be non-intrinsic (see Sect.~\ref{sec:biases}): for instance, we find that X-ray obscuration (Sect.~\ref{sec:biases}) and variability across the epochs or a low duty-cycle (Fig.~\ref{fig:fxfx} and Appendix~\ref{sec:moreplots}) are unlikely to be responsible for the almost 200 non-detected galaxies. Furthermore, the X-ray weakness was not found to depend on the variability significance in IR-selected galaxies (Fig.~\ref{fig:lxlo_irsign}), nor on the number of data points and total baseline in the optical light curves (bottom panel of Fig.~\ref{fig:lxlo_sigmavar}). We only find weak evidence that the stacked X-ray signal is slightly brighter for galaxies with higher significance variability in the optical (Appendix~\ref{sec:moreplots}), although no significant differences were found (see also Fig.~\ref{fig:lxlo_sigmavar}). Since, formally, our work was not able to confirm most MBH candidates despite the eRASS:4 survey being sensitive enough, another possibility we must conservatively consider is that variability-selected MBH samples are severely biased by unknown contaminants, or unknown methodological flaws, spread to all variability significance values. This would imply that these galaxies are inactive and that they lack significant accretion in their nuclei. Everything considered (see also Appendix~\ref{sec:moreplots}), this is admittedly very unlikely. Therefore, the observed X-ray weakness has to be intrinsic to the bulk of the low-mass galaxies population, or at the very least that selected via UVOIR variability. Hence, this might imply that a canonical X-ray corona is lacking in these nuclei. In Sect.~\ref{sec:discussion_xweak}, we discuss that a possible explanation for this might lie in the fundamental differences between the nuclei of low-mass galaxies and the more massive ones. For instance, the shallower potential well and clumpier interstellar medium in the former, compared to the latter. However, it remains to be quantitatively addressed whether these differences lead to a inefficient magnetization of the nuclear gas \citep[e.g.,][]{Begelman+2023:magn} and whether this ultimately affects the powering of the corona at very small scales ($\sim10$\,gravitational radii).
    
    An indirect way to confirm the presence of a systematic X-ray (and X-ray only) weakness in the MBHs SEDs, would be to analyze the UVOIR variability property \citep[e.g. with LSST;][]{Ivezic+2019:LSST} and radio incidence and X/radio ratios \citep[e.g. with ASKAP-EMU;][]{Norris+2011:EMU} of an X-ray selected MBH sample. If a comparably puzzling low confirmation rate is obtained, this would imply that all single-band searches are incomplete (and not only X-ray selections) and can not be used as representative for the MBH population. 
    Discouragingly, constraining the occupation fraction in low-mass galaxies was already known to be a challenging task in general \citep[e.g.,][]{Chadayammuri+2023:occfrac}. However, even if the bulk of the dwarf galaxy population were to be intrinsically X-ray weak, or with unusual SEDs, there is a minority of (observationally) well-behaved galaxies which are detected throughout the SED, providing useful lower limits for the active and occupation fractions \citep[e.g.,][]{Miller+2015:bhof,Gallo+2019:bhmf}. These would be less constraining than anticipated, but may still serve in ruling out pessimistic seeding models. Hence, this work serves as a pilot study for future synergies between eROSITA and LSST. We rely on the extensive simulated observations recently performed in \citet{Burke+2023:lsst} as benchmark for the expected number of variable MBHs detected by LSST. Following the assumptions and criteria used in \citet{Burke+2023:lsst}, we compare LSST predictions with our detection fractions between $M_* = 10^{8-10} M_{\astrosun}$ and below $z<0.055$: $3.4_{-1.0}^{+2.6}\%$ for eRASS1 and $6.4_{-1.5}^{+3.0}\%$ for eRASS:4. We adopt the predicted LSST MBHs numbers from \citet{Burke+2023:lsst} of $1.5_{-0.6}^{+0.6}\times 10^3$ and $5.9_{-1.1}^{+1.5}\times 10^3$, obtained using light and heavy seed models, respectively. Therefore, on the order of $\approx20-130$ and $\approx155-440$ in eRASS1 ($\approx45-195$ and $\approx235-695$ in eRASS:4), based on light and heavy seed models, of LSST's MBH candidates may be detected and, hence, confirmed. We note that these numbers are most likely lower limits, as LSST is expected to be more complete in sampling the intrinsic stellar mass and magnitude distribution, compared to the inhomogeneous sample used in this work (e.g. see Fig.~\ref{fig:hist} and~\ref{fig:detfrac}).

	\begin{acknowledgements}
    We thank the referee for their positive comments on the manuscript.
    
    R.A. is grateful to C. J. Burke, V. Baldassare, F. Pacucci, U. Chadayammuri, I. Chilingarian, M. Begelman, J. Silk and S. von Fellenberg for insightful discussions about their works.
 
	We acknowledge the use of the matplotlib package \citep{Hunter2007:matplotlib} and Photutils, an Astropy package for	detection and photometry of astronomical sources \citep{Bradley+2022:photutils1.4.0}.
    This research has made use of the SIMBAD database,
    operated at CDS, Strasbourg, France.
	
	R.A received support for this work by NASA through the NASA Einstein Fellowship grant No HF2-51499 awarded by the Space Telescope Science Institute, which is operated by the Association of Universities for Research in Astronomy, Inc., for NASA, under contract NAS5-26555. 

    G.P. acknowledges funding from the European Research Council (ERC) under the European Union’s Horizon 2020 research and innovation programme (Grant agreement No. [865637]), and support from Bando per il Finanziamento della Ricerca Fondamentale 2022 dell’Istituto Nazionale di Astrofisica (INAF): GO Large program.

	This work is based on data from eROSITA, the soft X-ray instrument aboard \emph{SRG}, a joint Russian-German science mission supported by the Russian Space Agency (Roskosmos), in the interests of the Russian Academy of Sciences represented by its Space Research Institute (IKI), and the Deutsches Zentrum für Luft- und Raumfahrt (DLR). The \emph{SRG} spacecraft was built by Lavochkin Association (NPOL) and its subcontractors, and is operated by NPOL with support from the Max Planck Institute for Extraterrestrial Physics (MPE). The development and construction of the eROSITA X-ray instrument was led by MPE, with contributions from the Dr. Karl Remeis Observatory Bamberg \& ECAP (FAU Erlangen-Nuernberg), the University of Hamburg Observatory, the Leibniz Institute for Astrophysics Potsdam (AIP), and the Institute for Astronomy and Astrophysics of the University of Tuebingen, with the support of DLR and the Max Planck Society. The Argelander Institute for Astronomy of the University of Bonn and the Ludwig Maximilians Universitaet Munich also participated in the science preparation for eROSITA. The eROSITA data shown here were processed using the eSASS software system developed by the German eROSITA consortium.
	
	The Legacy Surveys consist of three individual and complementary projects: the Dark Energy Camera Legacy Survey (DECaLS; Proposal ID \#2014B-0404; PIs: David Schlegel and Arjun Dey), the Beijing-Arizona Sky Survey (BASS; NOAO Prop. ID \#2015A-0801; PIs: Zhou Xu and Xiaohui Fan), and the Mayall z-band Legacy Survey (MzLS; Prop. ID \#2016A-0453; PI: Arjun Dey). DECaLS, BASS and MzLS together include data obtained, respectively, at the Blanco telescope, Cerro Tololo Inter-American Observatory, NSF’s NOIRLab; the Bok telescope, Steward Observatory, University of Arizona; and the Mayall telescope, Kitt Peak National Observatory, NOIRLab. Pipeline processing and analyzes of the data were supported by NOIRLab and the Lawrence Berkeley National Laboratory (LBNL). The Legacy Surveys project is honored to be permitted to conduct astronomical research on Iolkam Du’ag (Kitt Peak), a mountain with particular significance to the Tohono O’odham Nation. NOIRLab is operated by the Association of Universities for Research in Astronomy (AURA) under a cooperative agreement with the National Science Foundation. LBNL is managed by the Regents of the University of California under contract to the U.S. Department of Energy. This project used data obtained with the Dark Energy Camera (DECam), which was constructed by the Dark Energy Survey (DES) collaboration. Funding for the DES Projects has been provided by the U.S. Department of Energy, the U.S. National Science Foundation, the Ministry of Science and Education of Spain, the Science and Technology Facilities Council of the United Kingdom, the Higher Education Funding Council for England, the National Center for Supercomputing Applications at the University of Illinois at Urbana-Champaign, the Kavli Institute of Cosmological Physics at the University of Chicago, Center for Cosmology and Astro-Particle Physics at the Ohio State University, the Mitchell Institute for Fundamental Physics and Astronomy at Texas A\&M University, Financiadora de Estudos e Projetos, Fundacao Carlos Chagas Filho de Amparo, Financiadora de Estudos e Projetos, Fundacao Carlos Chagas Filho de Amparo a Pesquisa do Estado do Rio de Janeiro, Conselho Nacional de Desenvolvimento Cientifico e Tecnologico and the Ministerio da Ciencia, Tecnologia e Inovacao, the Deutsche Forschungsgemeinschaft and the Collaborating Institutions in the Dark Energy Survey. The Collaborating Institutions are Argonne National Laboratory, the University of California at Santa Cruz, the University of Cambridge, Centro de Investigaciones Energeticas, Medioambientales y Tecnologicas-Madrid, the University of Chicago, University College London, the DES-Brazil Consortium, the University of Edinburgh, the Eidgenossische Technische Hochschule (ETH) Zurich, Fermi National Accelerator Laboratory, the University of Illinois at Urbana-Champaign, the Institut de Ciencies de l’Espai (IEEC/CSIC), the Institut de Fisica d’Altes Energies, Lawrence Berkeley National Laboratory, the Ludwig Maximilians Universitat Munchen and the associated Excellence Cluster Universe, the University of Michigan, NSF’s NOIRLab, the University of Nottingham, the Ohio State University, the University of Pennsylvania, the University of Portsmouth, SLAC National Accelerator Laboratory, Stanford University, the University of Sussex, and Texas A\&M University. BASS is a key project of the Telescope Access Program (TAP), which has been funded by the National Astronomical Observatories of China, the Chinese Academy of Sciences (the Strategic Priority Research Program “The Emergence of Cosmological Structures” Grant \# XDB09000000), and the Special Fund for Astronomy from the Ministry of Finance. The BASS is also supported by the External Cooperation Program of Chinese Academy of Sciences (Grant \# 114A11KYSB20160057), and Chinese National Natural Science Foundation (Grant \# 12120101003, \# 11433005). The Legacy Survey team makes use of data products from the Near-Earth Object Wide-field Infrared Survey Explorer (NEOWISE), which is a project of the Jet Propulsion Laboratory/California Institute of Technology. NEOWISE is funded by the National Aeronautics and Space Administration. The Legacy Surveys imaging of the DESI footprint is supported by the Director, Office of Science, Office of High Energy Physics of the U.S. Department of Energy under Contract No. DE-AC02-05CH1123, by the National Energy Research Scientific Computing Center, a DOE Office of Science User Facility under the same contract; and by the U.S. National Science Foundation, Division of Astronomical Sciences under Contract No. AST-0950945 to NOAO.
	
	\end{acknowledgements}

	%
	%
	\bibliographystyle{aa} 
	\bibliography{bibliography} 
	
\begin{appendix}        
\section{Further diagnostics on the X-ray weakness of MBHs}
\label{sec:moreplots}	


\begin{figure}[h]
    \centering
    \includegraphics[width=0.99\columnwidth]{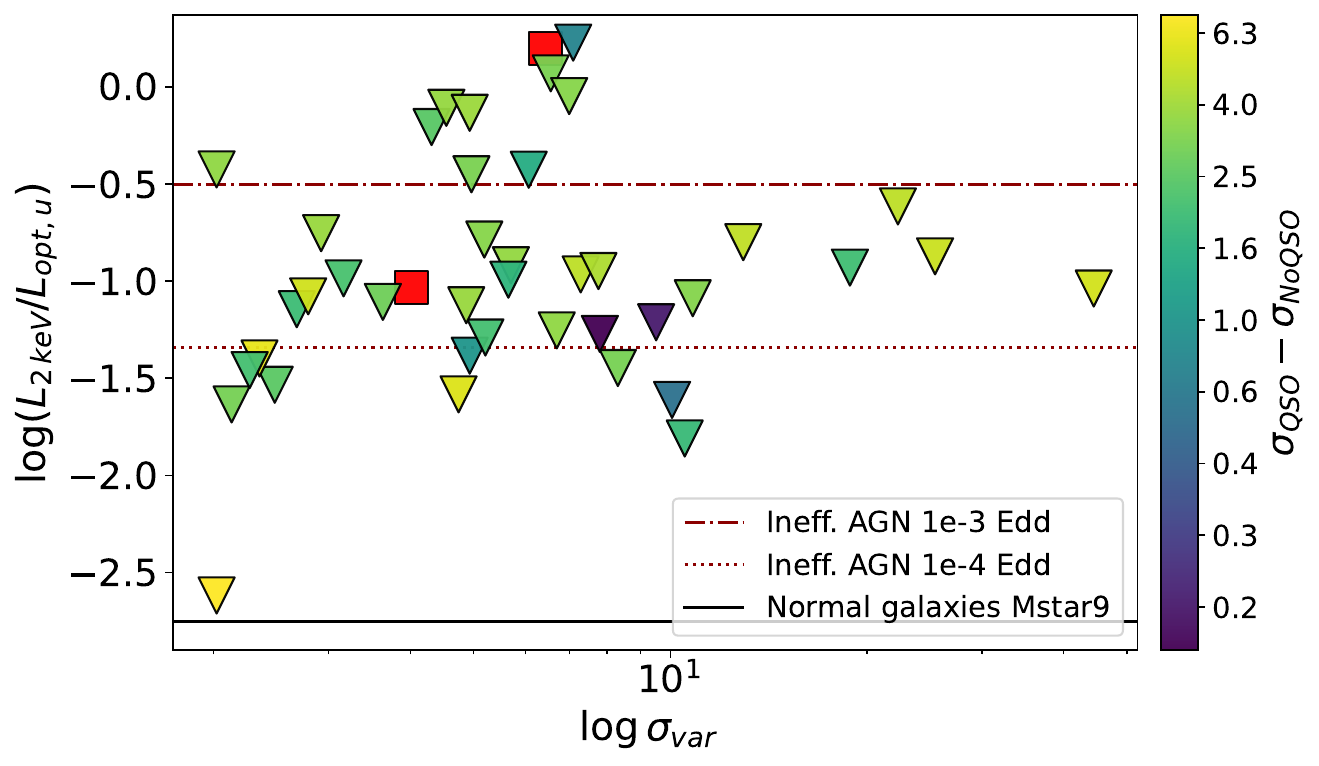}
    \includegraphics[width=0.99\columnwidth]{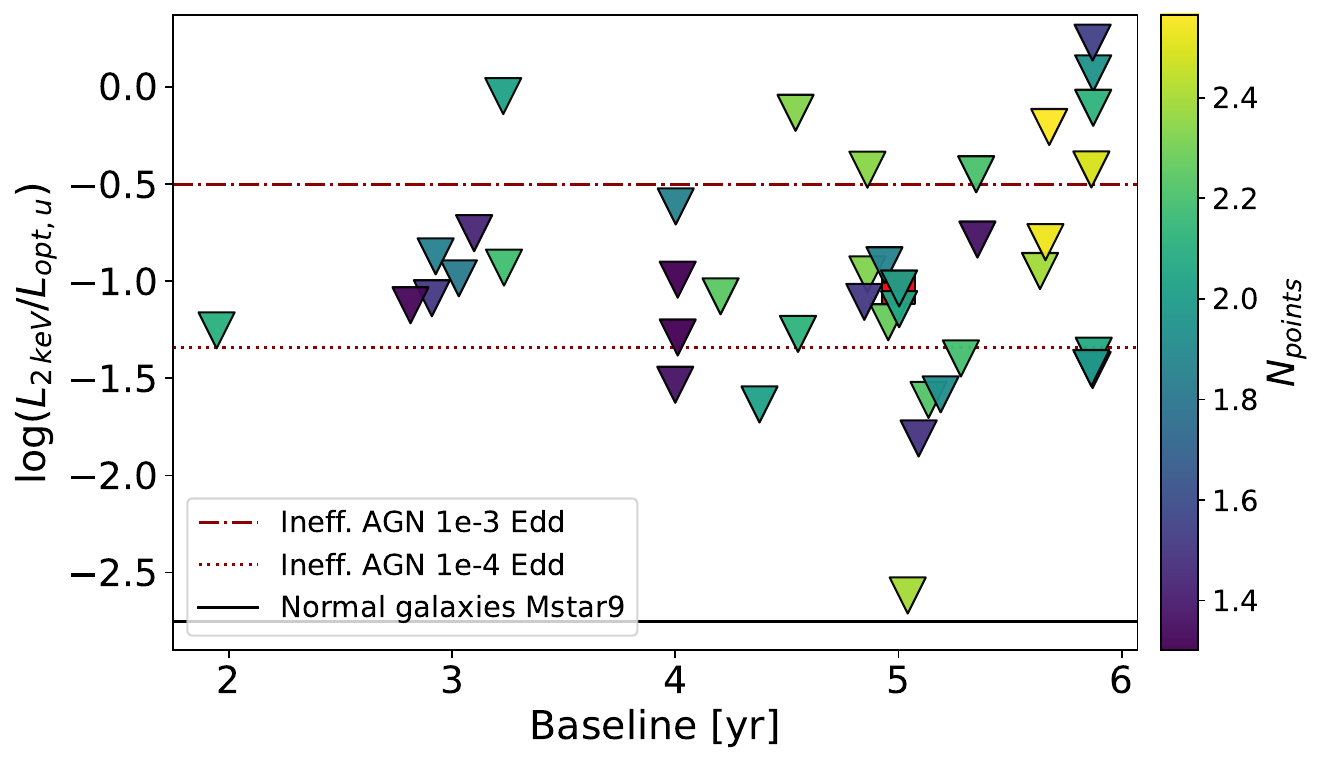}
    \caption{\emph{Top panel}: observed X-ray to optical ratio as in Fig.~\ref{fig:lxlo_mstar}, shown here as a function of optical variability significance and is color-coded as a function of the significance of AGN-like (compared to non-AGN-like) variability. Only non-X-ray-detected galaxies from \citet{Baldassare+2018:SDSS} and \citet{Baldassare+2020:PTF} are color-coded. X-ray detections are shown with red squares. \emph{Bottom panel}: same as the top panel, but as a function of baseline in years of the optical light curve, color-coded by the number of data points.}
    \label{fig:lxlo_sigmavar}
\end{figure}

\begin{figure}[h]
    \centering
    \includegraphics[width=0.99\columnwidth]{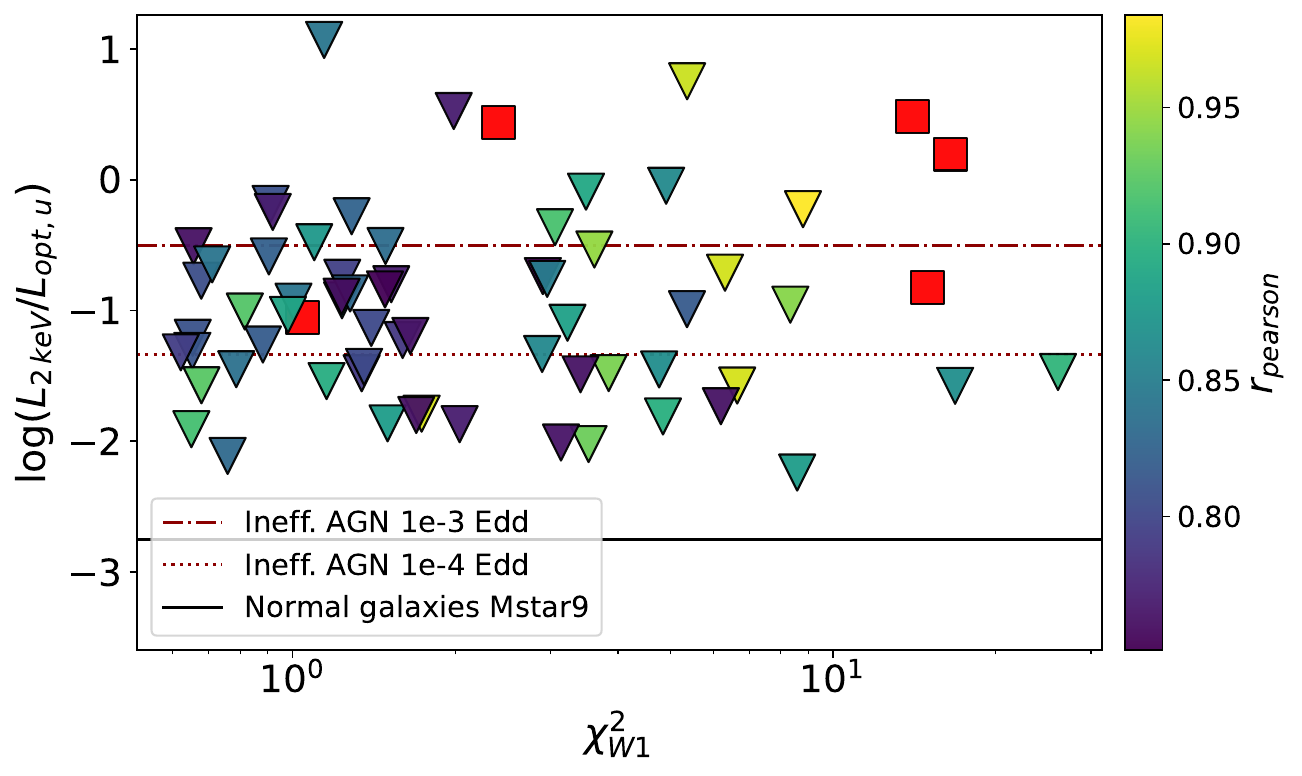}
    \caption{Same as Fig.~\ref{fig:lxlo_sigmavar}, but for the IR-selected galaxies from \citet{Ward+2022:ZTF_Wise}. X/O values are plotted as a function of $\chi_2$ in the W1 band light curve and color-coded by the Pearson correlation coefficient between W1 and W2 light curves.}
    \label{fig:lxlo_irsign}
\end{figure}

\begin{figure}[t]
    \centering
    \includegraphics[width=0.9\columnwidth]{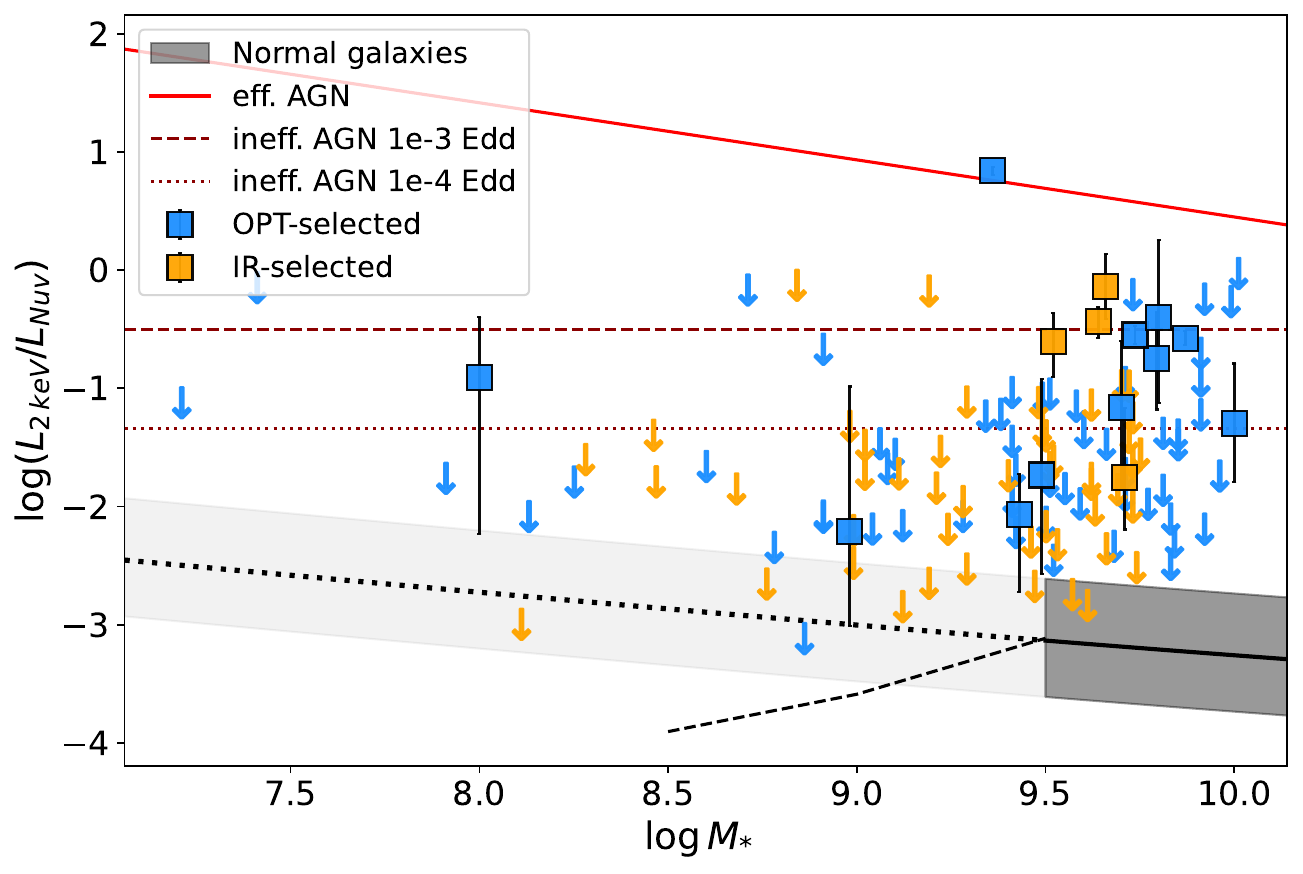}
    \caption{Same as Fig.~\ref{fig:lxlo_mstar}, but using the GALEX near-UV filter instead of the SDSS u-band filter.}
    \label{fig:lxlo_galex}
\end{figure}

Here, we perform some tests to further investigate the presence of biases in our interpretation of the systematic X-ray weakness observed in our sample. First, we check that X-ray weakness does not depend on the variability significance. We performed this test for the optically selected galaxies in \citet{Baldassare+2018:SDSS,Baldassare+2020:PTF}. In these works, the quantity $\sigma_{var}$ is the significance that the object is generally variable, while $\sigma_{QSO}$ that the damped random walk model adopted for AGN-like variability \citep{Kelly+2009:DRW} is significant compared to non-AGN-like variability, given by $\sigma_{NoQSO}$ \citep{Butler+2011:sigma}. These estimates yield high-purity in quasars samples \citep{Butler+2011:sigma} and we assume compatible purity is obtained for more nearby dwarf galaxies. Fig.~\ref{fig:lxlo_sigmavar} shows that the X-ray weak upper limits are not biased toward lower significance sources. Most X-ray weak upper limits have high $\sigma_{var}$ and $\sigma_{QSO} - \sigma_{NoQSO}$, therefore we do not expect that more than a handful of the parent MBHs in dwarf galaxies to be spuriously detected. To test this more quantitatively, we stacked the 39 galaxies within $\log M_*=9-10$ and below $z<0.1$, selected from from \citet{Baldassare+2020:PTF} and non-detected in eRASS:4. We divided low- and high-significance sources using $\sigma_{var}=6$ \citep{Baldassare+2020:PTF} as threshold, which grants an equal number of 20 and 19 galaxies in the two subsamples. Using as background estimate the median signal between $15-50\,$kpc (see Sect.~\ref{sec:xray}), the low-significance subsample is undetected in the stacked image with an upper limit at $L_{0.5-2.0\,keV}<4.2\times 10^{38}\,$erg\,s$^{-1}$. Conversely, the high-significance subsample is detected at $L_{0.5-2.0\,keV}=(9.3\pm7.2)\times 10^{38}\,$erg\,s$^{-1}$. However, if we use conservatively the 84th percentile of the signal between $15-50\,$kpc as background estimate (see Sect.~\ref{sec:xray}), the high-significance subsample is undetected as well, with an upper limit at $L_{0.5-2.0\,keV}<1.3\times 10^{39}\,$erg\,s$^{-1}$. Therefore, while this indicates that there is weak evidence of the high-significance subsample being brighter in X-rays, the difference is not significant enough. Finally, from the bottom panel of Fig.~\ref{fig:lxlo_sigmavar} we note that there are not obvious biases of having the deepest X-ray non-detections toward shorter baselines, or toward low number of data points, in the optical light curves. We perform the same check on the IR-selected galaxies from \citet{Ward+2022:ZTF_Wise}, where variability significance was expressed as a function of the Pearson correlation coefficient ($r_{pearson}$) between the binned W1 and W2 light curves and the related $\chi^2$ values (e.g. $\chi^2_{W1}$), both aimed to quantify variability compared to the median value of the light curve. Similarly to the optically selected sources, from Fig.~\ref{fig:lxlo_irsign} we note that the X-ray weak upper limits are not biased toward lower significance sources. Hence, we conclude that the spurious fraction in the parent sample of optically- and IR-variable galaxies is not significantly higher for lower-significance variability.


In Sect.~\ref{sec:discussion_xweak} and Fig.~\ref{fig:lxlo_mstar} we have inferred that the MBH population is X-ray weak compared to the X-ray flux predicted from the optical luminosity. Since the u-band filter used in Fig.~\ref{fig:lxlo_mstar} has an effective wavelength of $\sim3565\AA$, whilst the adopted scaling relations are calibrated at $\sim2500-3000\AA$ \citep{Arcodia2019:lxluv,Ruan+2019:aox}, we here test the use of the near-UV filter of GALEX \citep{Bianchi+2017:galex}, which has an effective wavelength of $\sim2300\AA$. We show the equivalent of Fig.~\ref{fig:lxlo_mstar}, but with GALEX data, in Fig.~\ref{fig:lxlo_galex}. We note that the comparison between observed X/O values and model predictions remains qualitatively the same and in fact using GALEX even fainter X/O values are obtained (cf. Fig.~\ref{fig:lxlo_mstar}).

\begin{figure}[t]
    \centering
    \includegraphics[width=0.99\columnwidth]{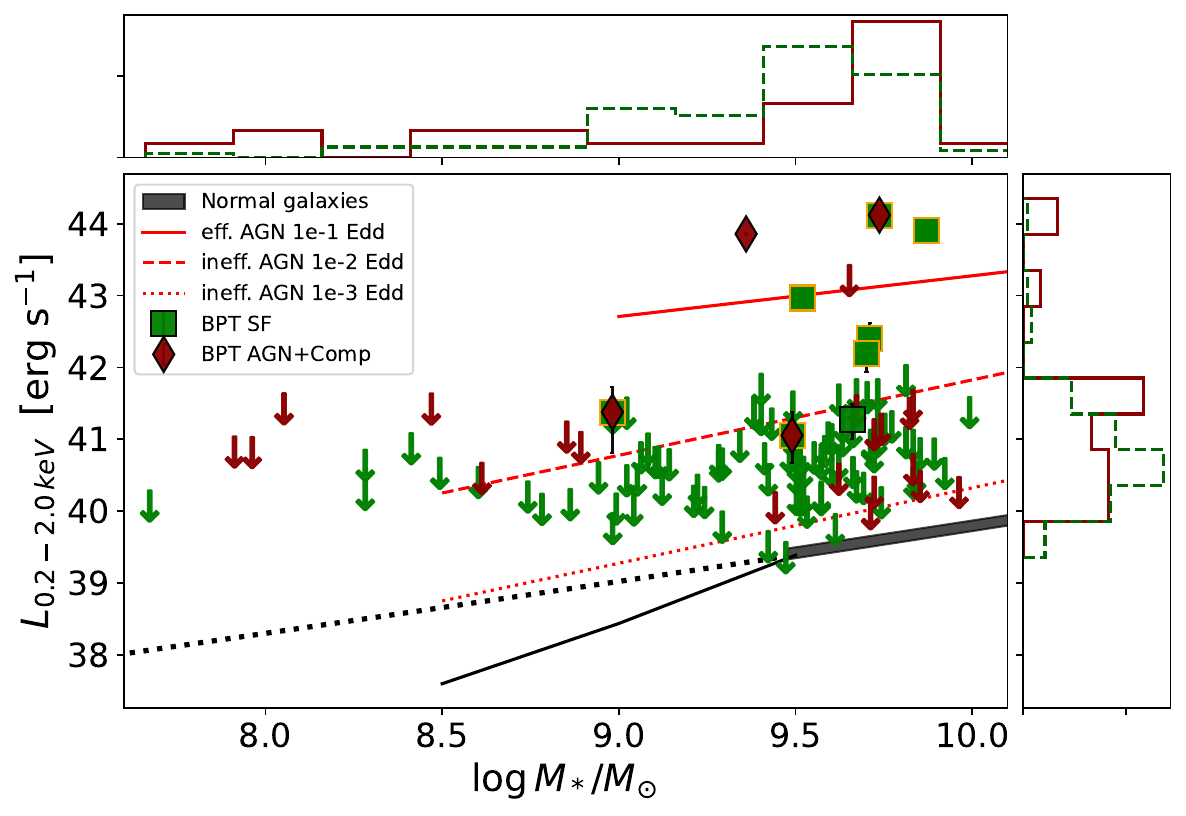}
    \includegraphics[width=0.99\columnwidth]{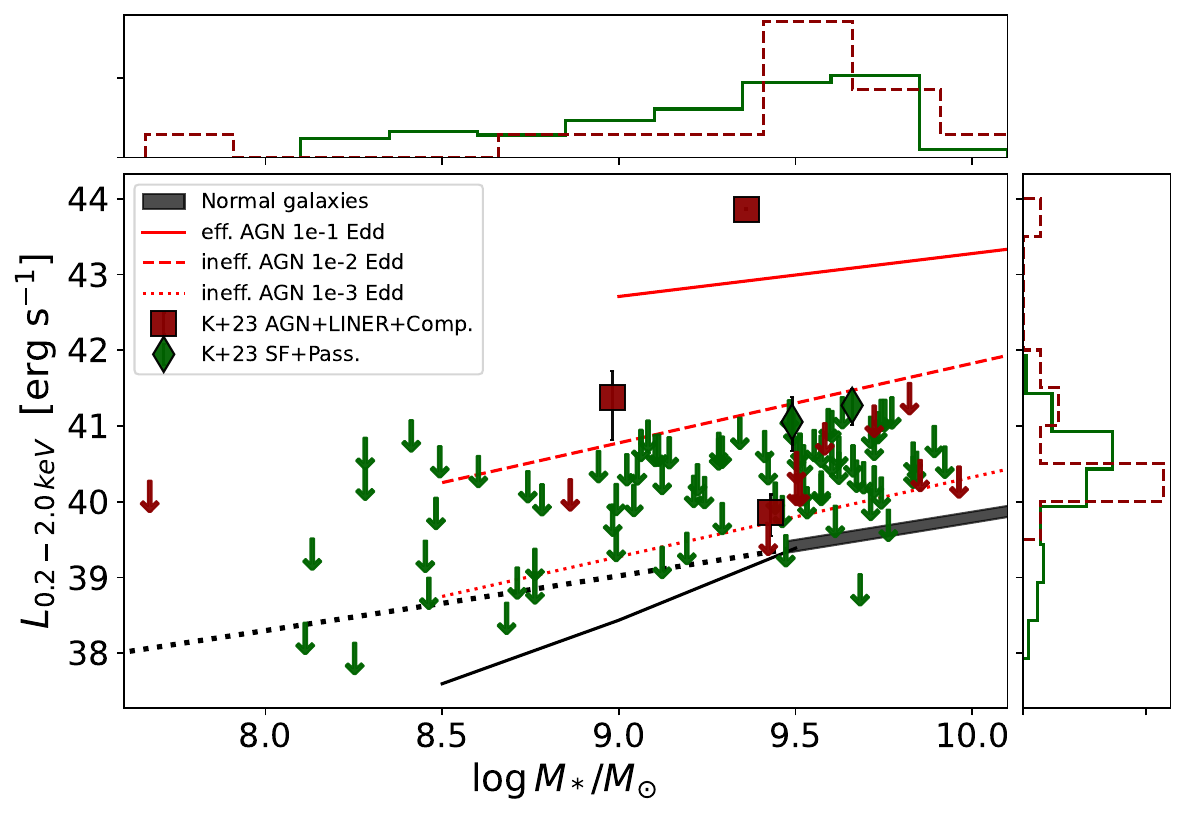}
    \includegraphics[width=0.95\columnwidth]{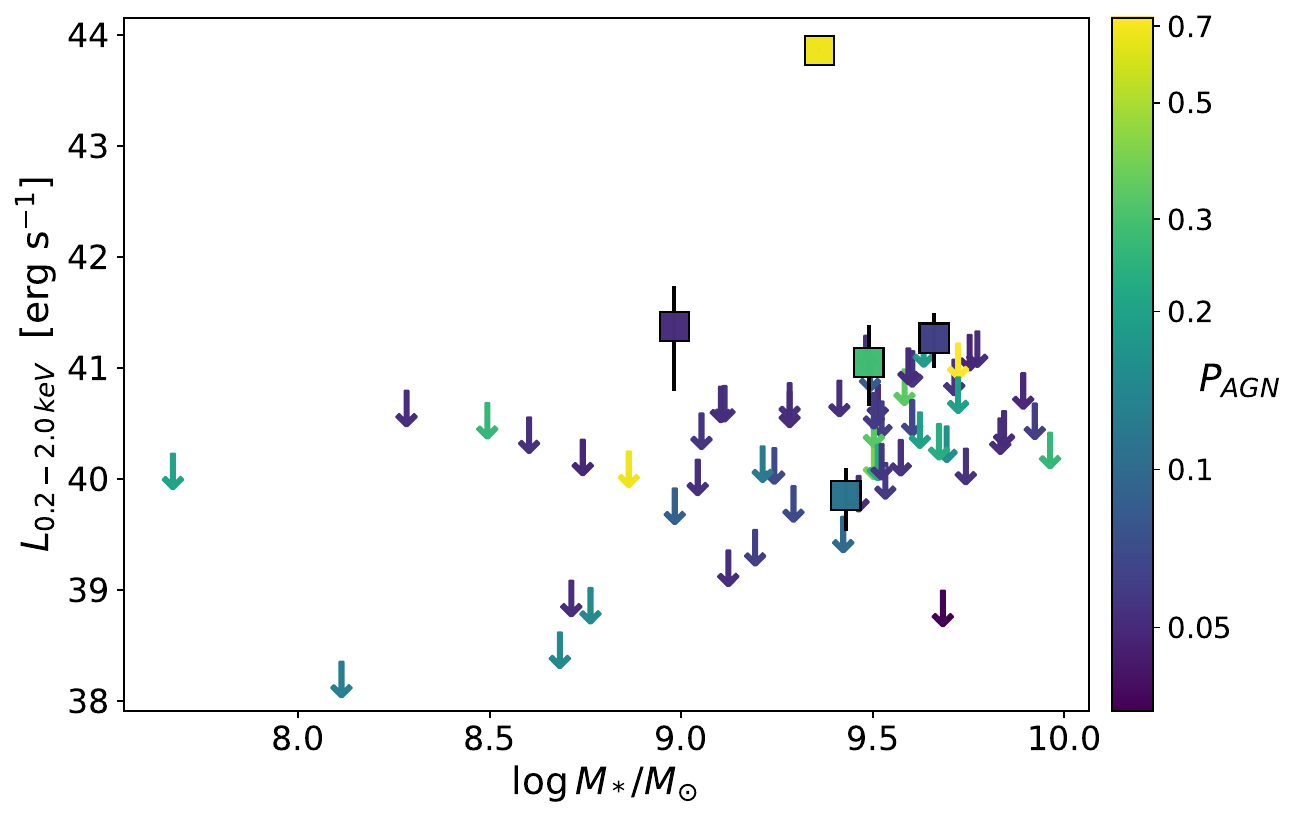}
    \caption{\emph{Top panel}: same as the bottom panel of Fig.~\ref{fig:lxmstar}, but color-coded as a function of BPT classification (green for star-forming galaxies and red for ``Composite'' and AGN) from RCSEDv2 (see text). Orange contours around X-ray detection of star-forming galaxies highlight sources with a broad $H\alpha$ component.
    \emph{Middle panel}: same as the top panel, but using photometric and spectroscopic classifications from the HECATE catalog (see text). In this subplot, we also show LINERs together with AGN and ``Composite'' galaxies (red), and the passive together with star-forming ones (green). \emph{Bottom panel}: same as the other panels, but galaxies are color-coded with a probabilistic estimate on the presence of an AGN, from photometric and spectroscopic classifications from the HECATE catalog (see text).
    }
    \label{fig:bpt}
\end{figure}

The bottom panel of Fig.~\ref{fig:lxmstar} and in Fig.~\ref{fig:lxlo_mstar} do not include the classification of the galaxies based on optical spectra. Here, we investigate the dependency of the observed X-ray weakness on the classification of the galaxy based on optical photometry and spectroscopy, as an independent proxy compared to the UVOIR selection. However, we note that the UVOIR variability method is knowingly selecting AGN candidates in galaxies classified as inactive \citep{Baldassare+2018:SDSS,Baldassare+2020:PTF}. First, we retrieved the galaxy classification of our sample from the Reference Catalog of galaxy SEDs \citep[RCSEDv2\footnote{\href{https://dev-rcsed2.voxastro.org}{\texttt{https://dev-rcsed2.voxastro.org}}};][]{Chilingarian+2012:rcsed,Chilingarian+2017:rcsed}, between $z=0.01-0.1$. The lower end is chosen to avoid aperture biases, the higher end is chosen to limit the analysis to the range in which X-ray non-detections were stacked. A handful of sources which were either missing in the database or had spectra with poor quality were excluded. This analysis was limited to 99 galaxies. We show in the top panel of Fig.~\ref{fig:bpt} the equivalent of the bottom panel of Fig.~\ref{fig:lxmstar}, to which we added subpanels with histograms and a different color coding. We highlight in green (squares for detections, arrows for non-detections) the galaxies classified as star-forming from the BPT narrow lines diagnostics \citep{Baldwin+1981:bpt}, whilst in red (diamonds and arrows) those classified as Composite or as AGN. In addition, we highlight with orange contours the galaxies classified as star-forming, but for which RCSED reports a significant detection of a broad $H\alpha$ line. 

Furthermore, we also estimate the activity classifications with the updated version of HECATE catalog (Kyritsis et al., in prep.). The classifications are based on two different methods. The first one is an advanced data-driven version of the traditional BPT diagrams, which utilizes a soft clustering scheme for classifying emission-line galaxies in different activity classes using simultaneously four emission-line ratios \citep{Stampoulis+2019:class}. The second one is based on the application of the Random Forest machine learning algorithm on mid/IR (W1-W2, W2-W3; WISE) and optical (g-r; SDSS) colors and can discriminate galaxies into 5 activity classes (i.e star-forming, AGN, ``Composite'', ``LINER'', and ``Passive''; \citealp{Daoutis+2023:class}).  Both activity classification methods are probabilistic, meaning that they provide the probability of a galaxy to belong in each class, and an example of their application is presented in the work of Kyritsis et al. (in prep.) for the selection of all the bona-fide star-forming galaxies which were observed by the eRASS1 all-sky survey. First, we confirmed that the two methods yielded similar results from one-another, compatibly with the top panel of Fig.~\ref{fig:bpt}. Then, in the middle panel of Fig.~\ref{fig:bpt} we show the combination of the two above-mentioned methods from the HECATE catalog: we represent a galaxy in red (green) if either the emission-lines diagnostics or the Random Forest consider it as AGN, ``LINER'' or ``Composite'' (star-forming or passive). Similarly to the top panel of the same figure, there is no significant different among the two sets of classifications. Finally, in the bottom panel of Fig.~\ref{fig:bpt} we color-code the plot with the probability of a galaxy to host an AGN based on the Random Forest algorithm. Again, we identify no major bias: x-ray detections are found at all $P_{AGN}$ and non-detections do not seem to strongly depend on $P_{AGN}$ either. This tests highlights that there is no significant difference between the X-ray weakness of galaxies classified as star-forming, compared to those classified as active.

\begin{figure}[t]
    \centering
    \includegraphics[width=0.99\columnwidth]{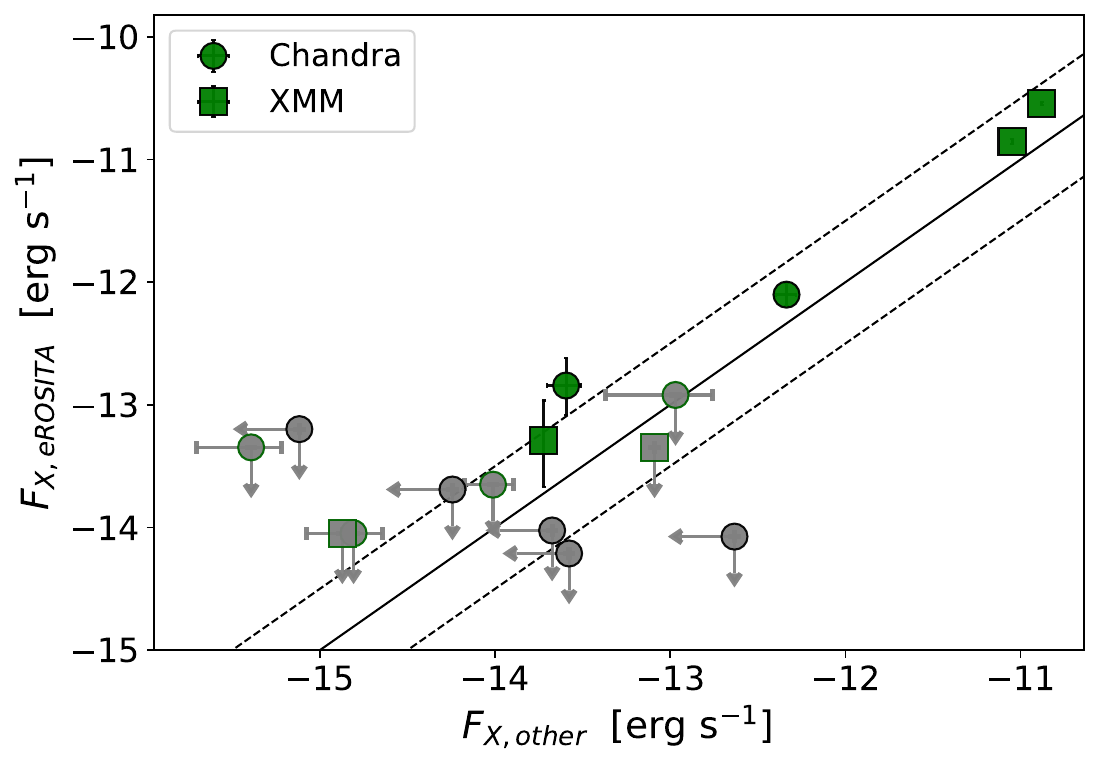}
    \caption{Comparison between eROSITA eRASS:4 X-ray fluxes and archival \emph{XMM-Newton} \citep{Webb+2020:xmmdr12} and \emph{Chandra} data \citep{Evans+2020:chandra,Messick+2023:ovar} of the same galaxy.
    }
    \label{fig:fxfx}
\end{figure}

Furthermore, we have checked the impact of X-ray variability, although it is expected to yield a scatter in both brighter and fainter directions and not the latter only. As a matter of fact, we have crossmatched the eRASS:4 low-z galaxies with the fourth \emph{XMM-Newton} serendipitous source catalog \citep{Webb+2020:xmmdr12} and the \emph{Chandra} Source Catalog \citep{Evans+2020:chandra}. We added a handful of sources from \citet{Messick+2023:ovar}, which were not included in the catalogs (namely NSA IDs 156688, 104881, 51928, 67333, 124477). We show in Fig.~\ref{fig:fxfx} the resulting comparison, where the 1:1 (with related 0.5\,dex scatter) is show with a solid (dashed) line. Different energy bands might have been used across different sources, although consistent bands are used between eROSITA and other missions for the individual source. Different symbols are used for \emph{XMM-Newton} (squares) and \emph{Chandra} (circles), while different colors highlight eROSITA detected (green) and non-detected (gray) sources. Detections with  \emph{XMM-Newton} and \emph{Chandra} are highlighted with green contours for visualization purpose. All the sources detected by eROSITA and either \emph{XMM-Newton} or \emph{Chandra} (with observations taken between 2015 and 2022) show compatible fluxes across the different epochs. All eROSITA upper limits (apart from one) are brighter than the detection with \emph{XMM-Newton} or \emph{Chandra}, therefore they are compatible with the 1:1 and were not supposed to be detected by eROSITA. Upper limits in both missions (gray data points with black contours) are, by definition, compatible with the 1:1. Therefore, we confirm that the impact of variability or a low duty cycle in these galaxies has to be minimal.

\begin{figure}[t]
    \centering
    \includegraphics[width=0.99\columnwidth]{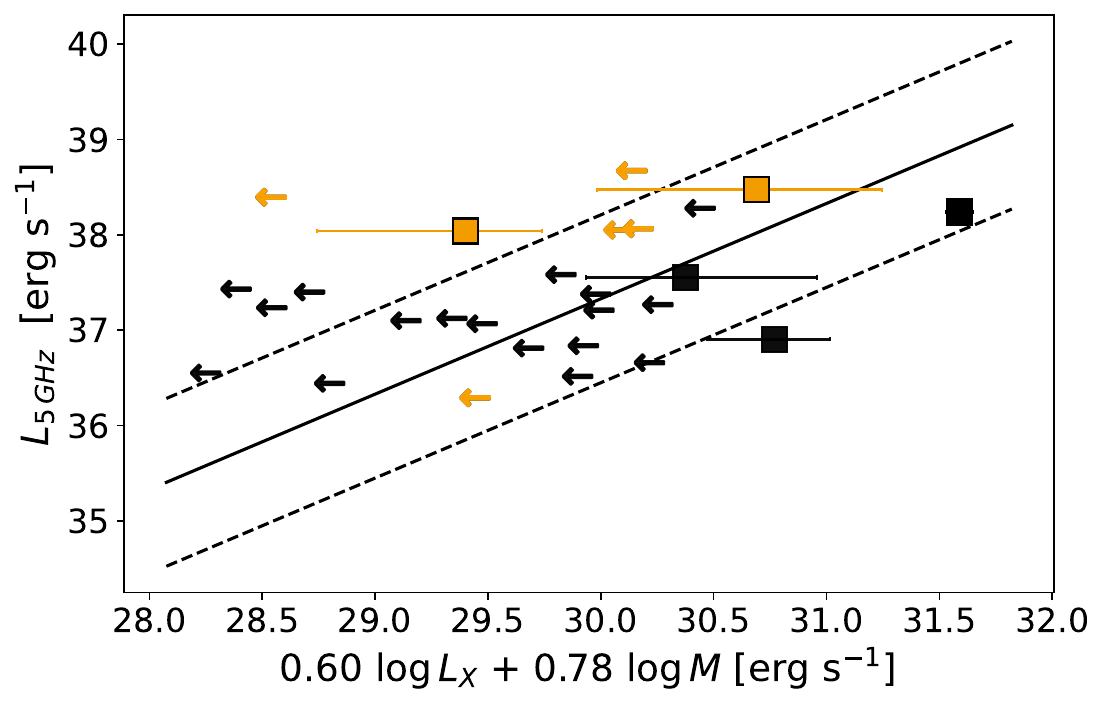}
    \includegraphics[width=0.99\columnwidth]{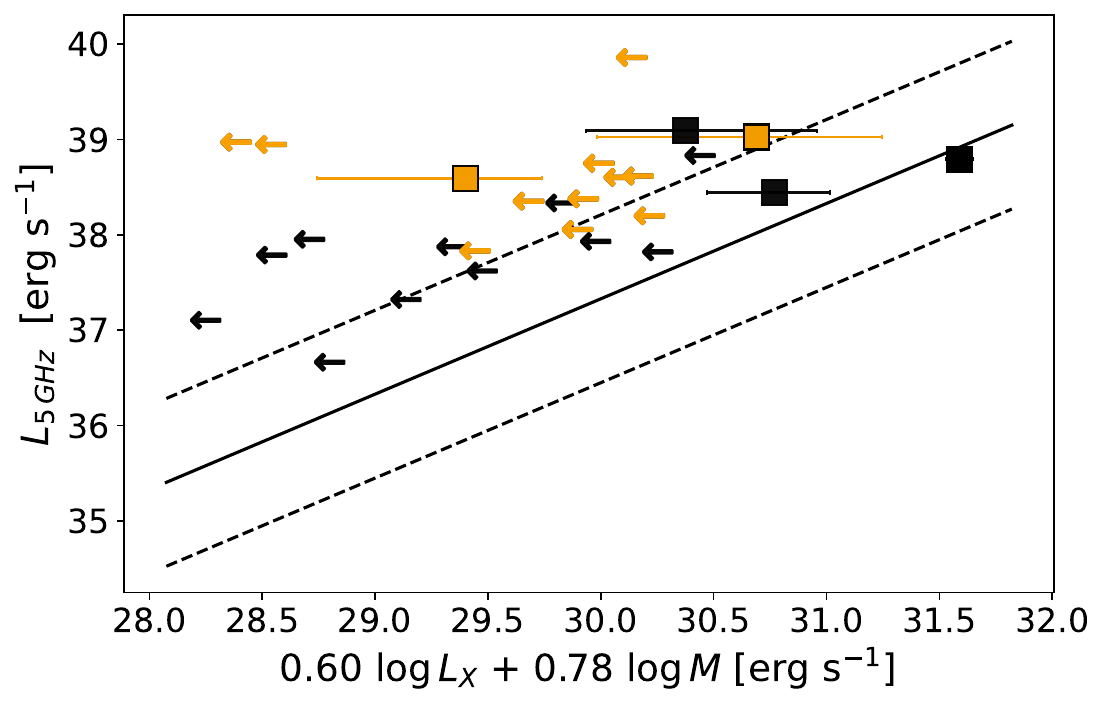}
    \caption{The fundamental plane of black hole accretion \citep{Merloni+2003:FP} is shown with the solid line, with its $\sim0.88\,$dex scatter. We show all the sources in our low-z sample which can be matched to an archival radio observation. We highlight in orange galaxies with a SFR estimate and a radio luminosity brighter than that predicted by SFR. The \emph{top panel} shows radio fluxes extrapolated assuming a radio spectral index of -1, the \emph{bottom} using a flat slope.
    }
    \label{fig:fp}
\end{figure}

In order to quantify how the X-ray weakness compares with the radio properties of the MBHs, we cross-matched our low-z sample (Fig.~\ref{fig:lxlo_mstar}) with radio archives\footnote{\href{https://heasarc.gsfc.nasa.gov/W3Browse/master-catalog/radio.html}{Link to radio catalog}}, the Rapid ASKAP Continuum Survey \citep{Mcconnell+2020:RACS,Hale+2021:racs} and the second data release of the LOFAR Two-metre Sky Survey \citep{Shimwell+2022:lofar}. We then convert the observed radio fluxes to $5\,$GHz luminosities assuming both a spectrum with radio spectral index -1 (top panel of Fig.~\ref{fig:fp}) and a flat spectrum (bottom panel of Fig.~\ref{fig:fp}). We estimated the black hole masses from the stellar masses of the galaxies \citep{Reines+2015:coev} and plotted our sources in the fundamental plane of black hole accretion \citep{Merloni+2003:FP}. We show this in Fig.~\ref{fig:fp}. To be conservative, we draw the main conclusions from the top panel as it shows the faintest 5\,GHz luminosity from the extrapolations. Realistically, radio spectra of these sources would be a mixed bag between slopes of minus one and zero, therefore between the two panels. We note that both X-ray and radio fluxes are likely contaminated by the galaxy. Therefore we computed the radio luminosity at $5\,$GHz as predicted by star-formation in the galaxy \citep{Ranalli+2003:sfr}. In Fig.~\ref{fig:fp}, we highlight in orange MBHs with a SFR estimate available from the MPA-JHU catalog \citep{Brinchmann+2004:mpajhu} and with a radio luminosity greater than that predicted for the galaxy alone \citep{Ranalli+2003:sfr}. The sample statistic is now very low, although X-ray weak $3\sigma$ upper limits remain. This is more evident if a flat radio spectrum is assumed. 
Hence, MBHs appear to be X-ray weaker even compared to the fundamental plane, including the large intrinsic scatter of $\sim0.88\,$dex of the relation. This is at odds with the interpretation that X-ray weakness is simply due to the low-luminosity nature of these MBHs. 

\section{Tables}

We present eRASS1 and eRASS:4 results on all MBH candidates in Table~\ref{tab:results} and more details on the detected sources in Table~\ref{tab:results_detections}.

	\onecolumn
		\begin{ThreePartTable}
			
			\LTcapwidth=\columnwidth
\begin{longtable}{ccccccccc}
				\caption{Input MBH candidates from optical/IR/UV variability and related eROSITA information from aperture photometry and spectroscopy.}
				\label{tab:results} \\
				
				\toprule
				\multicolumn{5}{c}{Input} &
				\multicolumn{2}{c}{eRASS1} &
				\multicolumn{2}{c}{eRASS:4}
				\\
				\multicolumn{1}{c}{RA} &
				\multicolumn{1}{c}{Dec} &
				\multicolumn{1}{c}{$z$} &
				\multicolumn{1}{c}{$M*$} &
				\multicolumn{1}{c}{Ref.\tnote{a}} &
				\multicolumn{1}{c}{$P_B$\tnote{b}} &
				\multicolumn{1}{c}{$L_{0.2-2.0\,keV}$\tnote{c}} &
				\multicolumn{1}{c}{$P_B$\tnote{b}} &
				\multicolumn{1}{c}{$L_{0.2-2.0\,keV}$\tnote{c}} \\
				\endfirsthead
				
				\multicolumn{3}{l}{\footnotesize Continued} \\
				\toprule
				\multicolumn{5}{c}{Input} &
				\multicolumn{2}{c}{eRASS1} &
				\multicolumn{2}{c}{eRASS:4}
				\\
				\multicolumn{1}{c}{RA} &
				\multicolumn{1}{c}{Dec} &
				\multicolumn{1}{c}{$z$} &
				\multicolumn{1}{c}{$M*$} &
				\multicolumn{1}{c}{Ref.\tnote{a}} &
				\multicolumn{1}{c}{$P_B$\tnote{b}} &
				\multicolumn{1}{c}{$L_{0.2-2.0\,keV}$\tnote{c}} &
				\multicolumn{1}{c}{$P_B$\tnote{b}} &
				\multicolumn{1}{c}{$L_{0.2-2.0\,keV}$\tnote{c}} \\
				
				\endhead
				\midrule
				\endfoot
				\bottomrule
				\endlastfoot

				\midrule
  				48.25895      &      -0.686379      &       0.131   &   9.8    &   Ba18        &   2.62e-32    &   $\mathbf{43.41_{43.29}^{43.53}\,(_{43.11}^{43.67})}$             &       6.19e-66      &     $\mathbf{43.19_{43.11}^{43.28}\,(_{43.0}^{43.39})}$  \\
  				49.42965      &      0.326904       &       0.069   &   9.7    &   Ba18        &   1.51e-13    &   $\mathbf{42.32_{42.15}^{42.49}\,(_{41.91}^{42.7})}$              &       3.23e-28      &     $\mathbf{42.19_{42.08}^{42.3}\,(_{41.94}^{42.46})}$  \\
  				47.61596      &      -0.830791      &       0.08    &   9.74   &   Ba18        &   0    &   $\mathbf{44.2_{44.18}^{44.22}\,(_{44.15}^{44.25})}$              &       0      &     $\mathbf{44.12_{44.11}^{44.14}\,(_{44.09}^{44.15})}$ \\
  				55.87608      &      -7.58542       &       0.036   &   9.84   &    Ba20       &   1           &   $<39.78\,(40.85)$                                                &       0.94          &     $<39.36\,(40.35)$                                    \\
  				154.78365     &      19.98218       &       0.039   &   9.88   &    Ba20       &   1           &   $<39.94\,(41.06)$                                                &       0.44          &     $<39.85\,(40.8)$                                     \\
  				57.34663      &      -11.99095      &       0.032   &   9.36   &    Ba20       &   0    &   $\mathbf{44.02_{44.01}^{44.02}\,(_{44.0}^{44.04})}$              &       0      &     $\mathbf{43.86_{43.85}^{43.86}\,(_{43.85}^{43.87})}$ \\
  				123.4507      &      37.83548       &       0.029   &   9.07   &    Ba20       &   1           &   $<39.88\,(40.98)$                                                &       0.23          &     $<40.1\,(40.87)$                                     \\
\end{longtable}
			Sources which were masked out (see Sect.~\ref{sec:xray}) are shown with dashes.
			\begin{tablenotes}
				\item[a] Reference for input coordinates, redshift and stellar mass $M_*$: Ba18 stands for \citet{Baldassare+2018:SDSS}; Ba20 for \citet{Baldassare+2020:PTF}; B22 for \citet{Burke+2022:DES}; W22Z for ZTF-selected sources from \citet{Ward+2022:ZTF_Wise}; Ki20 for \citet{Kimura+2020:HSC}; Sh22 for \citet{Shin+2022:intranight}; W22W for Wise-selected sources from \citet{Ward+2022:ZTF_Wise}; S20 for \citet{Secrest+2020:Wise}; Ha23 for \citet{Harish+2023:Wise}; Was22 for \citet{Wasleske+2022:Galex}.
				\item[b] No-source probability $P_B$ (Eq.~\ref{eq:Pb}). Sources are considered detected at $P_B<=0.0003$ (and are highlighted in bold).
				\item[c] Logarithmic X-ray luminosity in the rest-frame 0.2-2.0\,keV range, in units of $\log(\,$erg\,s$^{-1})$. For detected sources (in bold), median and 16th, 84th percentile values are shown first, with 1st and 99th in parenthesis. For non detected sources, 84th and 99th percentile values are shown as $1\sigma$ and $3\sigma$ upper limits, respectively. \\
			\end{tablenotes}
		\end{ThreePartTable}

\begin{table}[!h]
	\caption{eRASS:4 detections matched with \emph{XMM-Newton}, \emph{Chandra}, \emph{ROSAT} and \emph{Swift-XRT}. Sources with no previous X-ray detections are highlighted in bold.}
	\label{tab:results_detections}
	\centering
	\begin{ThreePartTable}
		\begin{tabular}{cccccccc}
			\toprule
			\multicolumn{1}{c}{RA} &
			\multicolumn{1}{c}{Dec} &
			\multicolumn{1}{c}{$L_{0.2-2.0\,keV}$} &
			\multicolumn{1}{c}{XMM} &
			\multicolumn{1}{c}{Chandra} &
			\multicolumn{1}{c}{ROSAT} &
			\multicolumn{1}{c}{Swift} &
			\multicolumn{1}{c}{Comments}
			\\
			\midrule
			\textbf{48.25895}      &      \textbf{-0.686379}              &     $\mathbf{43.19_{43.11}^{43.28}\,(_{43.0}^{43.39})}$      &  --  &    --   &  --   &  --    &                 \\
			\textbf{49.42965}      &      \textbf{0.326904}                 &     $\mathbf{42.19_{42.08}^{42.3}\,(_{41.94}^{42.46})}$      &  --  &    --   &  --   &  --    &                 \\
			47.61596      &      -0.830791                &     $44.12_{44.11}^{44.14}\,(_{44.09}^{44.15})$     &  Y   &    --   &   Y   &  --    &    AGN      \\
			57.34663      &      -11.99095                &     $43.86_{43.85}^{43.86}\,(_{43.85}^{43.87})$     &  Y   &    --   &   Y   &   Y  &    BL Lac \\
			191.64688     &      2.36911                  &     $43.91_{43.9}^{43.92}\,(_{43.89}^{43.93})$      &  Y   &    --   & Y &  Y    &   AGN     \\
			220.05293     &      2.79542                  &     $41.05_{40.91}^{41.19}\,(_{40.67}^{41.38})$     &  Y   &    Y    & Y & --     &    AGN             \\
			\textbf{184.28861}     &      \textbf{12.45432}                 &     $\mathbf{39.86_{39.74}^{39.96}\,(_{39.54}^{40.1})}$      &  --  &    --   &  --   &  --    &                 \\
			9.2741003     &      -44.66830                &     $43.88_{43.63}^{44.13}\,(_{43.02}^{44.54})$     &  Y   &    --   &  --   &  --    &   XMM serendipitous   \\
			52.512298     &      -27.54680                &     $44.05_{43.93}^{44.17}\,(_{43.77}^{44.32})$     &  Y   &    --   &   Y   &  --    &     AGN            \\
			132.857567    &      39.594941                &     $41.38_{41.19}^{41.54}\,(_{40.81}^{41.73})$     &  --  &    Y    &  --   & --    &  AGN               \\
			187.264621    &      29.779443                &     $42.4_{42.3}^{42.49}\,(_{42.16}^{42.62})$       &  --  &    Y    &  --   & --    &  AGN                \\
			\textbf{196.822679}    &      \textbf{13.646658}                &     $\mathbf{41.27_{41.16}^{41.37}\,(_{41.01}^{41.49})}$     &  --  &    --   &  --   & --    &                 \\
			148.661629    &      40.534554                &     $42.97_{42.93}^{43.01}\,(_{42.87}^{43.06})$     &  Y   &    Y    &   Y   & --    &   AGN              \\
			214.864575    &      4.753834                 &     $43.36_{43.31}^{43.41}\,(_{43.24}^{43.48})$     &  --  &    --   &   --  &  Y     &      BL Lac           \\
			160.860308    &      11.090063                &     $43.16_{43.14}^{43.18}\,(_{43.11}^{43.21})$     &  --  &    --   &   --  &   Y     &    AGN              \\	
            8.6218 & -43.3488                &     $43.94_{43.73}^{44.14}\,(_{43.42}^{44.41})$     &  Y &  Y   &  Y  &    Y    &          \\
            10.4139 & -43.7233                &     $43.90_{43.61}^{44.45}\,(_{43.07}^{44.45})$     & Y  &  Y   &  Y  &   Y     &     QSO $z\sim2$            \\

			\bottomrule
		\end{tabular}
	\end{ThreePartTable}
\end{table}
	
\twocolumn

\end{appendix}
	
\end{document}